\documentclass[a4paper,11pt]{article}
\usepackage{jheppub}

\pdfoutput=1
\usepackage{hyperref}
\usepackage{graphicx}
\usepackage[utf8]{inputenc}
\usepackage{xspace}
\usepackage{amsmath,amssymb}
\usepackage{bbm}
\usepackage{xcolor}
\usepackage{booktabs}
\usepackage{subcaption}
\usepackage{mathtools}
\usepackage[ruled,norelsize]{algorithm2e}
\usepackage{empheq}
\usepackage{cancel}
\usepackage{multirow}
\usepackage{slashed}
\usepackage{bbold}

\newcommand{\as}{\alpha_s}

\DeclareFontFamily{U}{rcjhbltx}{}
\DeclareFontShape{U}{rcjhbltx}{m}{n}{<->rcjhbltx}{}
\newcommand{\rd}{{d}}

\title{Quark and gluon two-loop beam functions for leading-jet
  \boldmath $p_T$ and slicing at NNLO}

\author[a,b]{Samuel Abreu,}
\author[c]{Jonathan R.~Gaunt,}
\author[a]{Pier Francesco Monni,}
\author[d]{Luca Rottoli,}
\author[e]{\\Robert Szafron}

\affiliation[a]{CERN, Theoretical Physics Department, CH-1211 Geneva
  23, Switzerland}
\affiliation[b]{Higgs Centre for Theoretical Physics, School of Physics and Astronomy,
 The University of Edinburgh, Edinburgh EH9 3FD, Scotland, United Kingdom}
\affiliation[c]{Department of
  Physics and Astronomy, University of Manchester, Manchester M13 9PL,
  United Kingdom}
\affiliation[d]{Department of Physics, University of Z\"urich, CH-8057 Z\"urich, Switzerland}
\affiliation[e]{Department of Physics, Brookhaven National Laboratory,
Upton, N.Y., 11973, U.S.A.}

\emailAdd{samuel.abreu@cern.ch}
\emailAdd{jonathan.gaunt@manchester.ac.uk}
\emailAdd{pier.monni@cern.ch}
\emailAdd{luca.rottoli@physik.uzh.ch}
\emailAdd{rszafron@bnl.gov}

\preprint{CERN-TH-2022-118, ZU-TH 30/22}
\abstract{
  We compute the complete set of two-loop beam functions for the
  transverse momentum distribution of the leading jet produced in
  association with an arbitrary colour-singlet system. Our results
  constitute the last missing ingredient for the calculation of the
  jet-vetoed cross section at small veto scales at the
  next-to-next-to-leading order, as well as an important ingredient
  for its resummation to next-to-next-to-next-to-leading logarithmic
  order.
  Our calculation is performed in the soft-collinear effective theory
  framework with a suitable regularisation of the rapidity divergences
  occurring in the phase-space integrals. We discuss the occurrence of
  soft-collinear mixing terms that might violate the factorisation
  theorem, and demonstrate that they are naturally absorbed into the
  beam functions at two loops in the exponential rapidity
  regularisation scheme when performing a multipole expansion of the
  measurement function.
  As in our recent computation of the two-loop soft function, we
  present the results as a Laurent expansion in the jet radius $R$. We
  provide analytic expressions for all flavour channels in $x$ space
  with the exception of a set of $R$-independent non-logarithmic terms
  that are given as numerical grids. We also perform a fully numerical
  calculation with exact $R$ dependence, and find that it agrees with
  our analytic expansion at the permyriad level or better.
  Our calculation allows us to define a next-to-next-to-leading order
  slicing method using the leading-jet $p_T$ as a slicing variable. As
  a check of our results, we carry out a calculation of the Higgs
  and $Z$ boson total production cross sections at the next-to-next-to-leading
  order in QCD.}

\begin{document}
\setlength{\parskip}{0pt}
\maketitle
\flushbottom


\section{Introduction}
\label{sec:intro}

Precision measurements at hadron colliders often rely on jet vetoes to
reduce the impact of background due to QCD radiation.
This is done by rejecting events containing jets with a transverse
momentum exceeding some cutoff value $p_T^{\rm veto}$ that is often
much smaller than the large momentum transfer of the hard scattering
process $Q$. This strategy finds common applications in the field of
Higgs physics, as well as in a number of electro-weak and QCD
measurements at the Large Hadron Collider (LHC), and has therefore
motivated a large number of
studies~\cite{Banfi:2012yh,Becher:2012qa,Banfi:2012jm,Banfi:2013eda,Becher:2013xia,Stewart:2013faa,Banfi:2015pju,Becher:2014aya,Moult:2014pja,Jaiswal:2014yba,Monni:2014zra,Wang:2015mvz,Dawson:2016ysj,Monni:2019yyr,Kallweit:2020gva}.
For instance, the state-of-the-art prediction for the jet-vetoed Higgs
production cross section involves the resummation of the large
logarithms $\ln (p_T^{\rm veto}/Q)$ up to next-to-next-to-leading
logarithmic (NNLL) order
\cite{Banfi:2012jm,Becher:2013xia,Stewart:2013faa}. These references
also include non-logarithmic terms relative to the Born at
${\cal O}(\alpha_s^2)$ (often referred to as NNLL$^\prime$),
numerically extracted from fixed-order calculations.
The above results are matched to fixed order calculations up to
next-to-next-to-next-to-leading order
(N$^3$LO)~\cite{Banfi:2015pju}.
The above numerical extractions of the non-logarithmic corrections
are, to the best of our knowledge, not entirely publicly available
and only encode information about the final convolution of such
non-logarithmic corrections with the parton densities. Conversely,
in order to access their perturbative dependence on the
longitudinal momentum fraction prior to the convolution, a
dedicated computation becomes necessary.

In this article, we directly compute the complete set of
${\cal O}(\alpha_s^2)$ non-logarithmic terms at high precision as
a function of the jet radius $R$, making them readily available
for the resummation of the jet-vetoed cross-section at
NNLL$^\prime$ for all color singlet production processes.
They are also a critical ingredient for the N$^3$LL resummation,
with the only missing ingredient being the three-loop rapidity
anomalous dimension.
This level of theoretical accuracy is demanded by the outstanding
experimental precision foreseen at the LHC in the coming years.
Specifically, we consider the calculation of the two-loop beam
functions entering the factorisation and resummation of the jet-vetoed
cross section.
Beam functions describe the dynamics of radiation collinear to the
beam direction in high-energy hadron collisions.
Together with our recent calculation of the two-loop soft function in
Ref.~\cite{Abreu:2022sdc}, the results presented here can be used to
efficiently calculate the leading singular terms of the jet-vetoed
cross section at small values of the veto scale (i.e., up to power
corrections in $p_T^{\rm veto}/Q$) at the next-to-next-to-leading
order (NNLO).
This importantly allows us to construct a slicing method for NNLO
calculations in QCD using the leading-jet transverse momentum as a
slicing variable.

We work in the framework of soft-collinear effective field theory
(SCET)~\cite{Bauer:2000yr,Bauer:2001yt,Bauer:2002nz,Beneke:2002ph,Beneke:2002ni}.
More specifically, the jet-veto cross section belongs to the class of
SCET$_{\rm II}$ problems, which are affected by the so-called
factorisation (or collinear) anomaly~\cite{BenekeFA}, connected to the
presence of rapidity divergences~\cite{Collins:1992tv,BenekeFA} in the
ingredients of the factorisation theorem. Such divergences are not
regulated by the standard dimensional regularisation scheme and
therefore an additional (rapidity) regulator must be introduced. Here
we use the exponential regularisation scheme~\cite{Li:2016axz},
consistently with our recent calculation of the two-loop soft
function~\cite{Abreu:2022sdc}.

The validity of SCET factorisation for this observable at arbitrary
logarithmic order has been the subject of debate in the
literature~\cite{Tackmann:2012bt,Stewart:2013faa,Becher:2013xia}.
Indeed, particular attention must be paid to the presence of
${\cal O}(R^2)$ soft-collinear mixing terms that might violate the
factorisation theorem for this observable.
In particular, Ref.~\cite{Becher:2013xia} presents an argument as to
why such terms should cancel if one performs a consistent multipole
expansion of the measurement functions. On the hand,
Refs.~\cite{Tackmann:2012bt,Stewart:2013faa} suggest that such
${\cal O}(R^2)$ terms are not captured by the conventional SCET
factorisation structure and therefore it is unclear how to perform
their resummation beyond NNLL order.
Following the observation of Ref.~\cite{Becher:2013xia}, we explicitly
show here that within the exponential regularisation scheme such a
multipole expansion of the jet-clustering measurement function leads
to the cancellation of the factorisation breaking terms, in the regime
in which the jet radius $R$ is treated as an $\mathcal{O}(1)$
parameter.\footnote{The resummation of small-$R$ logarithms in the regime $R\ll 1$ was performed in Ref.~\cite{Banfi:2015pju} for Higgs production and found to have a small numerical impact.}
This explicitly confirms the validity of the factorisation
theorem at NNLO, and constitutes an important step towards the
resummation at the N$^3$LL order.

The article is organised as follows.
In section~\ref{sec:factorisation}, we review the factorisation theorem
for the production of a colour-singlet with a veto on the transverse
momentum of the leading jet and we discuss the definition of the beam
functions in the presence of a rapidity
regulator. Section~\ref{sec:computation} contains a discussion of our
analytic computation of the two-loop beam functions as a small-$R$
expansion, as well as the details related to the zero-bin subtraction and the
cancellation of the soft-collinear mixing terms. 
Section \ref{sec:numeric} reviews our numerical calculation of the beam 
functions that retains the full-$R$ dependence, and compares the results 
obtained by the analytic and numerical computations, 
finding good agreement between the two.
Finally, in section~\ref{sec:slicing} we construct a phase-space slicing
scheme based on the leading-jet transverse momentum for fully
differential NNLO calculations of the production of colour-singlet
systems. We test the scheme, and our results, by calculating the NNLO
total cross section for Higgs and $Z$ boson production at the LHC.
Finally, our conclusions are given in section \ref{sec:conclusions}.

\section{Factorisation of leading-jet transverse momentum in SCET}
\label{sec:factorisation}
We begin by recalling the factorisation theorem for the jet-vetoed
cross-section. We consider the production of an arbitrary
colour-singlet system of total invariant mass $Q$ in proton-proton
collisions. The cross section, differential in the system's kinematics
$\rd\Phi_{\rm Born}$ and with a veto on the transverse momentum of the
leading jet $p_{T}^{\rm jet} < p_{T}^{\rm veto}$, factorises in the
limit $p_{T}^{\rm veto}\ll Q$ as
($\rd\sigma(p_{T}^{\rm veto})\equiv \frac{\rd\sigma(p_{T}^{\rm
    veto})}{\rd\Phi_{\rm
    Born}}$)~\cite{Tackmann:2012bt,Becher:2012qa,Stewart:2013faa,Becher:2013xia}
\begin{align}
\label{eq:factorisation}
\rd\sigma(p_{T}^{\rm veto}) &\equiv
                                                    \sum_{F=q,g}
                                                     |{\cal A}^F_{\rm
                                                     Born}|^2\,{\cal
                                                     H}^F(Q;\mu)\notag\\
  &\times \,{\cal B}^F_n(x_1,Q,p_{T}^{\rm veto},R^2;\mu,\nu) \,{\cal B}^F_{\bar n}(x_2,Q,
  p_{T}^{\rm veto},R^2;\mu,\nu)\,{\cal S}^F(p_{T}^{\rm veto},R^2;\mu,\nu)\,,
\end{align}
where ${\cal A}^F_{\rm Born}$ is the Born amplitude for the production
of the colour-singlet system, and $\mu$ and $\nu$ denote the
renormalisation and rapidity scales, respectively.
The index $F$ indicates the flavour configuration of the initial
state, i.e.~either $q\bar{q}$ ($F=q$) or $gg$ ($F=g$), and for
simplicity we will drop it from now on when referring to the
ingredients of the factorisation theorem~\eqref{eq:factorisation}.
The hard function ${\cal H}$ describes the dynamics at the hard scale,
i.e.~with virtuality $\mu\sim Q$. This scale is integrated-out in the
SCET construction. Therefore, the hard function contains purely
virtual contributions, and it is defined as the squared matching
coefficient of the leading-power two-jet SCET current, i.e.
\begin{equation}
{\cal H}^F(Q;\mu) = |{\cal C}^F(Q;\mu)|^2\,.
\end{equation}
The soft function ${\cal S}$ describes the dynamics of soft radiation
off the initial-state partons and is defined as a matrix element of
soft Wilson lines.
Lastly, the beam functions are denoted by ${\cal B}_n$ and
${\cal B}_{\bar n}$. Their two-loop calculation is the main focus of
this paper. They are defined by matrix elements of collinear fields in
SCET and describe the \mbox{(anti-)collinear} dynamics of radiation
along the light-cone directions $n^\mu$ and ${\bar n}^{\mu}$ of the
beams.

Within the SCET formalism, the resummation of the logarithms
$\ln p_{T}^{\rm veto}/Q$ appearing in~Eq.~\eqref{eq:factorisation} is
achieved by evolving each of the functions in the factorisation
theorem from their canonical scales to two common $\mu$, $\nu$ scales.
The hard matching coefficient obeys the renormalisation group equation (RGE)
\begin{equation}
\frac{d}{d\ln\mu}  \ln {\cal C}^F (Q;\mu)= \Gamma_{\rm cusp}^F(\alpha_s(\mu))
\ln\frac{-Q^2}{\mu^2} + \gamma_{H}^F(\alpha_s(\mu))\,,
\end{equation}
where $\Gamma_{\rm cusp}^F$ and $\gamma_{H}^F$ are the cusp and hard
anomalous dimensions of the quark ($F=q$) or gluon ($F=g$) form factors,
renormalised in the $\overline{\rm MS}$ scheme.
The hard function's canonical scale is the hard scale $\mu=Q$.
The beam functions ${\cal B}_i$ depend, in addition to the
renormalisation scale $\mu$, on the rapidity regularisation scale
$\nu$. They satisfy a system of coupled evolution equations (see
e.g.~Refs.~\cite{Becher:2013xia,Stewart:2013faa}). The RGE is given by
\begin{equation}
\frac{d}{d\ln\mu}\ln {\cal B}^F_i(x,Q,p_{T}^{\rm veto},R^2;\mu,\nu) = 2\,\Gamma_{\rm cusp}^F(\alpha_s(\mu))
\ln\frac{\nu}{Q} + \gamma_{B}^F(\alpha_s(\mu))\,,
\end{equation}
and  the rapidity evolution equation reads
\begin{equation}\label{eq:rapEvB}
\frac{d}{d\ln\nu}\ln {\cal B}^F_i(x,Q,p_{T}^{\rm veto},R^2;\mu,\nu) = 2
\int_{p_{T}^{\rm veto}}^{\mu}\frac{d\mu^\prime}{\mu^\prime}\Gamma_{\rm
  cusp}^F(\alpha_s(\mu^\prime)) -\frac{1}{2} \gamma_\nu^F(p_{T}^{\rm veto},R^2)\,,
\end{equation}
where $\gamma_\nu^F$ denotes the observable-dependent rapidity
anomalous dimension.
The boundary condition for the $\{\mu,\nu\}$  evolution is set at
the canonical scales $\mu=p_{T}^{\rm veto}$ and $\nu=Q$.
Finally, the evolution equations for the soft function read
\begin{align}
\frac{d}{d\ln\mu}\ln {\cal S}^F (p_{T}^{\rm veto},R^2;\mu,\nu) =& 4\,\Gamma_{\rm cusp}^F(\alpha_s(\mu))
  \ln\frac{\mu}{\nu} + \gamma_{S}^F(\alpha_s(\mu))\,,\notag\\
\frac{d}{d\ln\nu}\ln {\cal S}^F (p_{T}^{\rm veto},R^2;\mu,\nu) =&
-4\int_{p_{T}^{\rm veto}}^{\mu}\frac{d\mu^\prime}{\mu^\prime}\Gamma_{\rm
  cusp}^F(\alpha_s(\mu^\prime)) + \gamma_\nu^F(p_{T}^{\rm veto},R^2)\,,  
\end{align}
with canonical scales $\mu=\nu=p_{T}^{\rm veto}$.
The dependence on the rapidity anomalous dimension cancels between the
evolution of the soft and beam functions in the framework of the
rapidity renormalisation group~\cite{Chiu:2011qc,Chiu:2012ir}.  The
soft ($\gamma_{S}^F$) and collinear ($\gamma_{B}^F$) anomalous
dimensions are related to the hard anomalous dimension $\gamma_{H}^F$
by the invariance of the physical cross section under a change of the
renormalisation scale, that is
\begin{equation}
\label{eq:ADs}
2\gamma_{H}^F+\gamma_{S}^F+2 \gamma_{B}^F=0\,.
\end{equation}

The resummation of the jet-vetoed cross section at N$^k$LL requires
the cusp anomalous dimension $\Gamma_{\rm cusp}^F$ up to $k+1$ loops,
and the anomalous dimensions $\gamma_{H}^F$, $\gamma_{S}^F$,
$\gamma_{B}^F$, $\gamma_{\nu}^F$ up to $k$ loops.  The boundary
conditions (non-logarithmic terms) of the evolution equations need to
be known up to $k-1$ loops.
Achieving $\rm N^3LL$ accuracy for the jet-vetoed cross section
requires the knowledge of the non-logarithmic terms in
${\cal C}(Q;\mu)$ at two loops, which is given by the QCD on-shell
form-factor~\cite{Becher:2006nr} and has been known for a long time
\cite{Matsuura:1987wt,Gehrmann:2005pd}.
The two loop computation of the ${\cal S}$ function was presented in
our earlier article \cite{Abreu:2022sdc}.
Below, we focus on the evaluation of the two-loop beam
functions. While the two-loop anomalous dimensions are known, the
non-logarithmic terms are computed here for the first time.
Moreover, since the anomalous dimensions featuring in
Eq.~\eqref{eq:ADs} are known up to three
loops~\cite{Moch:2004pa,Gehrmann:2010ue,Li:2014afw}, with the results
presented in this work, the only missing ingredient for the N$^3$LL
computation is the three-loop rapidity anomalous dimension
$\gamma_{\nu}^F$.

\subsection{The beam functions}
The quark and gluon beam functions are defined as matrix elements of
non-local collinear operators between the proton states
$\left | P(p) \right\rangle$ carrying large momentum $p$. Specifically, we
have~\cite{Tackmann:2012bt,Becher:2012qa,Stewart:2013faa,Becher:2013xia}
\begin{align}
\label{eq:beam_def}
{\cal B}_n^q(x,Q,p_{T}^{\rm veto},R^2;\mu,\nu) &= 
 \frac{1}{2\pi} \int dt e^{-i x t \bar n\cdot p } \big\langle P(p) \big| 
 \overline{\chi}_n (t \bar n) \,\frac{\slashed{\bar{n}}}{2}\,{\mathbb M}(p_{T}^{\rm veto},R^2)\, \chi_n (0)
 \big | P(p) \big\rangle\,, \\
{\cal B}_n^g(x,Q,p_{T}^{\rm veto},R^2;\mu,\nu) &= 
- \frac{x \bar n \cdot p}{2\pi} \int dt e^{-i x t \bar n\cdot p } \big\langle P(p) \big| 
 \mathcal{A}^{\mu,a}_{\perp} (t \bar n) \,{\mathbb M}(p_{T}^{\rm veto},R^2)\,  \mathcal{A}_{\perp, \mu}^{a} (0) 
 \big | P(p) \big\rangle\,,
 \nonumber
\end{align}
where the collinear-gauge invariant collinear building blocks are
defined in terms of the fields
\begin{equation}
\chi(x) = W^\dagger \xi(x), \qquad \mathcal{A}^{\mu,A}_\perp =2 {\rm tr} \left[ W^\dagger \left[ i D^\mu_\perp W\right](x)  t^A \right]\,.
\end{equation}
Here $\xi(x)$ is the collinear quark field and $D^\mu_\perp$ is the
covariant $\perp$ collinear derivative. Gauge invariance is achieved
by introducing the collinear Wilson line $W(x)$ defined as
\begin{align}
W(x)  = \mathcal{P} \exp\left[i g\int_{-\infty}^0 ds \,\bar n A(x+ s \bar n )\right] \,.
\end{align}
The operator ${\mathbb M}(p_{T}^{\rm veto},R^2)$ acts on a given state
of $X_c$ collinear particles $|X_c\rangle$ by applying a veto on the
final-state jets of radius $R$ such that
$p_{T}^{\rm jet_i} < p_{T}^{\rm veto}$
\begin{align}
{\mathbb M}(p_{T}^{\rm veto},R^2) |X_c\rangle = \mathcal{M}(p_{T}^{\rm veto},R^2) |X_c\rangle\,,
\end{align}
with the phase space constraint $\mathcal{M}(p_{T}^{\rm veto},R^2)$ being
\begin{equation}
\mathcal{M}(p_{T}^{\rm veto},R^2) = \Theta(p_{T}^{\rm veto}-\max\{p_T^{\rm jet_i}\})
\Theta_{\rm cluster}(R^2)\,.
\end{equation}
Here $\max\{p_T^{\rm jet_i}\}$ is the transverse momentum of the
hardest jet, where jets are defined in the $E$ recombination
scheme~\cite{Cacciari:2011ma}. The constraint
$\Theta_{\rm cluster}(R^2)$ is the generic clustering condition on the
$X_c$ collinear final state particles, defined for a $k_T$-class of
jet algorithms~\cite{Cacciari:2008gp} with jet distance measures
\begin{equation}
  \label{eq:measures}
d_{ij} = \min\{k_{\perp i}^{2 p},k_{\perp j}^{2p}\} \left[(\Delta\eta_{i j})^2 +
(\Delta\phi_{i j})^2\right]\,,\quad d_{iB}=k_{\perp i}^{2 p}\,.
\end{equation}
Specific choices of the parameter $p$ correspond to the
anti-$k_T$~\cite{Cacciari:2008gp} algorithm ($p=-1$), the
Cambridge-Aachen~\cite{Dokshitzer:1997in,Wobisch:1998wt} algorithm
($p=0$), and the $k_T$~\cite{Catani:1993hr} algorithm ($p=1$). The
results obtained in this article are valid for any of these choices.
In Eq.~\eqref{eq:measures}, $k_{\perp i}$ is the transverse momentum
of particle $i$ with respect to the beam direction, and
$\Delta\eta_{i j}$ and $\Delta\phi_{i j}$ are the relative rapidity and
azimuthal angle between particles $i$ and $j$, respectively.
The particles are clustered sequentially with respect to the above
distance measure, as specified by the clustering condition
$\Theta_{\rm cluster}(R^2)$.

Since the definition~(\ref{eq:beam_def}) involves matrix elements
between proton states, the beam functions are in general
non-perturbative objects. However, for
$p_T^{\rm veto} \gg \Lambda_{\rm QCD}$, it is possible to
perturbatively match them onto the standard parton distribution
functions (PDFs) $ f_{F/P}(x,\mu)$ as
follows~\cite{Tackmann:2012bt,Becher:2012qa,Stewart:2013faa,Becher:2013xia},
\begin{align}\label{eq:beam-factroization}
{\cal B}_F(x,Q,p_{T}^{\rm veto},R^2;\mu,\nu) = \sum_{F'} \int_{x}^1 \frac{dz}{z} I_{FF'}(z,Q,p_{T}^{\rm veto},R^2;\mu,\nu) f_{F'/P}(x/z,\mu) + \mathcal{O}\left(\frac{\Lambda_{\rm QCD}}{p_T^{\rm veto}}\right)\,,
\end{align}
where the standard proton PDFs are defined as
\begin{align}
\label{eq:pdf_def}
f_{q/P}(x,\mu) &= 
 \frac{1}{2\pi} \int dt e^{-i x t \bar n\cdot p } \big\langle P(p) \big| 
 \overline{\chi}_n (t \bar n) \,\frac{\slashed{\bar{n}}}{2}\, \chi_n (0)
 \big | P(p) \big\rangle\,, \\
f_{g/P}(x,\mu) &= 
- \frac{x \bar n \cdot p}{2\pi} \int dt e^{-i x t \bar n\cdot p } \big\langle P(p) \big| 
 \mathcal{A}^{\mu,a}_{\perp} (t \bar n) \,  \mathcal{A}_{\perp, \mu}^{a} (0) 
 \big | P(p) \big\rangle\,,
 \nonumber
\end{align}
for the quark and the gluon, respectively.

The perturbative matching kernels
$I_{FF'}(z,Q,p_{T}^{\rm veto},R^2;\mu,\nu)$ are short-distance Wilson
coefficients, whose computation is the focus of this article. At the
LO, the collinear partons do not radiate and we have
$I_{FF'}(z,Q,p_{T}^{\rm veto},R^2;\mu,\nu)= \delta_{FF'}\delta(1-z)$.
The computation of radiative corrections to this relation up to the
two-loop order will be the subject of
Section~\ref{sec:computation}.\sloppy

The formal definition of the matching in
Eq.~\eqref{eq:beam-factroization} requires splitting the generic
collinear fields into the perturbative collinear modes with the
virtuality of the order of $p_T^{\rm veto}$ and the PDF-collinear
modes with virtuality $\Lambda_{\rm QCD}$. The non-perturbative PDFs
are then defined in terms of the matrix elements of the PDF-collinear
modes only, while the perturbative collinear modes are integrated out
from the theory. This distinction can often
be ignored in practice at the leading power in SCET.
Hence, in the rest of this article, we will
refer to both types of collinear fields as simply collinear modes.

Even though Eq.~(\ref{eq:beam-factroization}) is written as an
identity relating specific matrix elements, it represents an
operatorial identity, which does not depend on the specific choice of
the external states. Thus, to compute the matching kernels we can
replace the external proton states by perturbative partonic
states. With this replacement, the bare partonic PDF for finding a
parton $i$ inside parton $j$ becomes
\begin{align}
f^{\rm bare}_{i/j}(x) = \delta_{ij}\delta(1-x)\;,
\end{align}
which is valid to all orders in perturbation theory and the
partonic matrix elements in Eq.~(\ref{eq:beam_def}) are then directly equal
to the bare perturbative marching kernels.
For this reason, in what follows, we will refer to the $I_{FF^\prime}$
coefficients interchangeably as matching coefficients or beam
functions.


\section{Analytic computation of the quark and gluon beam functions}
\label{sec:computation}
In this section, we discuss the computation of the renormalised
matching coefficients $I_{FF'}(x,Q,p_T^{\rm veto},R^2;\mu,\nu)$,
obtained after the renormalisation of the collinear PDFs and of the
remaining UV divergences.
The perturbative expansion of the matching kernels in powers of the
strong coupling constant $\alpha_s$ is defined as
\begin{align}
I_{FF'} = \sum_{k=0}^{\infty} \left(\frac{\alpha_s}{4\pi}\right)^k I^{(k)}_{FF'} \;.
\end{align}
To efficiently compute the matching coefficients
$I_{FF'}(x,Q,p_T^{\rm veto},R^2;\mu,\nu)$, we decompose each beam
function into the sum of the beam function for a specifically chosen
reference observable and a remainder term. 
The reference observable is chosen so that it has the same 
single-emission limit. With this choice, the two-loop
matching coefficients of the reference observable have the same 
divergences in the dimensional regularisation parameter around $d=4$ 
as those for leading-jet $p_T$, and the
remainder term can then be computed directly in four dimensions.
As in our calculation of the soft functions \cite{Abreu:2022sdc},
we take the transverse momentum of the colour singlet system
as the reference observable. The associated
beam functions are denoted
by $I^\perp_{FF'}(x,Q,p_T^{\rm veto};\mu,\nu)$. These are known up to
$\mathcal{O}(\alpha_s^3)$~\cite{Catani:2011kr,Catani:2012qa,Gehrmann:2014yya,Echevarria:2016scs,
  Luo:2019hmp,Luo:2019bmw,Luo:2019szz,Ebert:2020yqt} and we consider
the renormalised ${\cal O}(\alpha_s^2)$ result of
Refs.~\cite{Luo:2019hmp,Luo:2019bmw} as a reference in our calculation
as these are also computed within the exponential rapidity 
regularisation scheme. The remainder term
$\Delta I_{FF'}(x,Q,p_T^{\rm veto},R^2;\mu,\nu)$ accounts for the
effects of the jet clustering algorithm. The perturbative matching
coefficients are then expressed as
\begin{align}\label{eq:deffdiff}
I_{FF'}(x,Q,p_T^{\rm veto},R^2;\mu,\nu) = I^\perp_{FF'}(x,Q,p_T^{\rm veto};\mu,\nu)+\Delta I_{FF'}(x,Q,p_T^{\rm veto},R^2;\mu,\nu)\; .
\end{align}
The functions $I^\perp_{FF'}(x,Q,p_T^{\rm veto};\mu,\nu)$ are obtained
from the beam functions of the transverse-momentum of the colour
singlet system~\cite{Luo:2019hmp,Luo:2019bmw}, and we include the
one-loop and two-loop contributions in our ancillary
files.\footnote{To be precise, following Appendix~\ref{app:renorm}, we
  first perform the inverse Fourier transform of the renormalised beam
  functions obtained in Refs.~\cite{Luo:2019hmp,Luo:2019bmw}, which
  are defined in impact parameter space. Then we integrate them up to
  $p_T^{\rm veto}$, obtaining
  $I^\perp_{FF'}(x,Q,p_T^{\rm veto};\mu,\nu)$.}

We stress here that the decomposition~\eqref{eq:deffdiff} is simply a
convenient way of organising the calculation, and the ingredient
$I^\perp_{FF'}$ has different physical properties from the actual beam
functions entering transverse momentum resummation. A first difference
is that the latter are defined in impact parameter space, since they
are sensitive to the vectorial nature of transverse momentum
factorisation. A second, related difference concerns the gluon beam
functions, which in transverse momentum resummation receive a
correction from different Lorentz structures including a linearly
polarised contribution (see
e.g.~\cite{Catani:2010pd,Gehrmann:2014yya,Gutierrez-Reyes:2019rug,Luo:2019bmw}). This
leads to peculiar azimuthal correlations between radiation collinear
to the two initial state (beam) legs. The above effect is absent in
the gluon beam functions defined in Eq.~\eqref{eq:beam_def}, which are
already integrated over the azimuthal angle of the emitted
radiation. A simple physical explanation for this observation is 
that, unlike in the transverse momentum case, the jet algorithms
considered in the factorisation theorem~\eqref{eq:factorisation} will
never cluster together emissions collinear to opposite incoming legs,
therefore leaving no phase space for the azimuthal correlations to
occur.

The term $\Delta I_{FF'}(x,Q,p_T^{\rm veto},R^2;\mu,\nu)$ defined in
(\ref{eq:deffdiff}) contributes only when two or more real emissions
are present and consequently $\Delta I_{FF'}$ starts at
$\mathcal{O}(\alpha_s^2)$. At this order it can be computed
directly in $d=4$ space-time dimensions and only real
emission diagrams contribute, with the measurement function
\begin{align}\label{eq:DeltaJ}
\Delta {\cal M}(p_{T}^{\rm veto},R^2) \equiv \Theta(p_{T}^{\rm veto}-\max\{p_T^{\rm jet_i}\})
\Theta_{\rm cluster}(R^2)- \Theta\left(p_{T}^{\rm
  veto}-\left|\sum_{i\in X_c}\vec{k}_{\perp i}\right|\right)\,.
\end{align}

To set-up the calculation, we need to take into account that the
phase-space integrals, as in a typical SCET$_{\rm II}$ problem,
exhibit rapidity divergences which require additional regularisation.
A consistent computation of the soft and beam functions requires that
the same regulator is used in both calculations.  We therefore adopt
the exponential regulator defined in Ref.~\cite{Li:2016axz}, which
we also used in our recent calculation of the two-loop soft
function~\cite{Abreu:2022sdc}. This regularisation procedure is
defined by altering the phase-space integration measure for real
emissions, such that
\begin{equation}
  \label{eq:regulator_def}
\prod_{i} d^d k_i \delta(k^2_i)\theta(k^0_i)\to  \prod_{i} d^d k_i \delta(k^2_i)\theta(k^0_i)\exp\left[ \frac{-e^{-\gamma_E}}{\nu}(n \cdot k_i + \bar{n} \cdot k_i)\right]\,,
\end{equation}
where $\nu$ is the rapidity regularisation scale discussed in
section~\ref{sec:factorisation}.  The regularised beam functions are
obtained by performing a Laurent expansion about $\nu \to +\infty$ and
neglecting terms of $\mathcal{O}(\nu^{-1})$.
In the rest of this section, we outline the main technical aspects of
the calculation of the matching coefficients and present our results.

\subsection{The renormalised one-loop beam functions}
At $\mathcal{O}(\alpha_s)$, the jet algorithm does not play a role and
the correction term in Eq.~\eqref{eq:deffdiff} vanishes
\begin{align}
\Delta \mathcal{M}(p_T^{\rm veto}, R^2)=0.
\end{align}
After the renormalisation of the collinear PDFs and of the remaining
UV divergences, the one-loop result reads (see
e.g.~\cite{Luo:2019hmp,Luo:2019bmw})
\begin{align}
I_{qq}^{(1)}(x,Q,p_T;\mu,\nu) &=2C_F L_\perp (4 L_Q+3)\delta(1-x)- 4 L_\perp  P_{qq}^{(0)}+2 C_F (1-x),  \\
I_{qg}^{(1)}(x,Q,p_T;\mu,\nu) &=- 4 L_\perp P_{qg}^{(0)} + 4 T_F x(1-x),  \\
I_{gq}^{(1)}(x,Q,p_T;\mu,\nu) &=-4 L_\perp P_{gq}^{(0)} + 2 C_F x,  \\
I_{gg}^{(1)}(x,Q,p_T;\mu,\nu) &=8C_A L_\perp  L_Q\delta(1-x)- 4
                                L_\perp P_{gg}^{(0)} + 8 L_\perp \delta(1-x) \beta_0,
\end{align}
with $L_\perp = \ln \frac{\mu}{p^{\rm veto}_T}$,
$L_Q = \ln \frac{\nu}{p^-}$, and where the space-like splitting functions are
\begin{align}
P_{qq}^{(0)} &= C_F \left(\frac{1+x^2}{1-x}\right)_+, \\
P_{qg}^{(0)} &= T_F \left(x^2+(1-x)^2\right), \\
P_{gq}^{(0)} &= C_F \left(\frac{1+(1-x)^2}{x}\right), \\
P_{gg}^{(0)} &= 2 C_A  \left[\frac{x}{(1-x)_+}
               +\frac{1-x}{x}+x(1-x)\right] + 2 \beta_0\,\delta(1-x) \,,
\end{align}
with $\beta_0 = \frac{11}{3} C_A - \frac{4}{3} T_F n_F$.

\subsection{The renormalised two-loop beam functions}
\label{sec:2loopbeams}
In this section we outline the procedure used in our computation of
the two-loop beam functions.
In particular, we do not discuss further the calculation of the
already known $I^{\perp}_{FF'}(x,Q,p_T^{\rm veto},R^2;\mu,\nu)$
component in Eq.~\eqref{eq:deffdiff}, whose expression, after the
renormalisation of the collinear PDFs and of the UV divergences, can be
found in Refs.~\cite{Luo:2019hmp,Luo:2019bmw}.
The following discussion refers to the correction
$\Delta I_{FF^\prime}$ for a generic flavour channel, although
specific channels may present a simpler structure, and some of the
contributions discussed below may vanish in their calculation.

The momenta of the two real particles (either gluons or quarks) are
denoted by $k_i$, with $i=1,2$, and we adopt the following parametrisation for the
phase space on the r.h.s.~of Eq.~\eqref{eq:regulator_def}:
\begin{align}
k_i ^\mu= k_{i\perp} \left(\cosh\eta_i\,,\cos\phi_i\,, \sin\phi_i, \sinh \eta_i \,\right),\quad i=1,2\,,
\end{align} 
in terms of the (pseudo-)rapidities $\eta_i$, the azimuthal angles
$\phi_i$, and the magnitudes of  transverse momenta
$k_{i\perp}\equiv |\vec{k}_{i\perp}|$.
We then perform a change of variables in the parametrisation
of $k_2$,
\begin{align}\label{eq:coordinates}
\left\{k_{2\perp},\eta_2,\phi_2\right\}\rightarrow
  \left\{\zeta\equiv
  k_{2\perp}/k_{1\perp},\eta\equiv\eta_1-\eta_2,\phi
  \equiv \phi_1 - \phi_2\right\}\,,
\end{align}
in order to express its kinematics relative to that of $k_1$. With
this change of variables, the measurement function~\eqref{eq:DeltaJ}
takes the simple form
\begin{align}\label{eq:DeltaJ-2L}
  \begin{split}
\Delta {\cal M}(p_{T}^{\rm veto},R^2) \equiv &\left[\Theta(p_{T}^{\rm veto}-k_{1\perp}\max\{1,\zeta\})
- \Theta\left(p_{T}^{\rm
                                      veto}-k_{1\perp}\sqrt{1+\zeta^2+2\zeta\cos\phi}\right)\right]\\
                                & \times \Theta(\eta^2+\phi^2 - R^2)\,,
\end{split}
\end{align}
where we used the explicit form of $\Theta_{\rm cluster}(R^2)$ in the
variables defined above, namely the relation
\begin{align}\begin{split}
  \Theta(p_{T}^{\rm veto}-\max\{p_T^{\rm jet_i}\})
\Theta_{\rm cluster}(R^2)\equiv
\Theta(p_{T}^{\rm veto}-k_{1\perp}\max\{1,\zeta\})\Theta(\eta^2+\phi^2 - R^2)\\
+\Theta\left(p_{T}^{\rm veto}-k_{1\perp}\sqrt{1+\zeta^2+2\zeta\cos\phi}\right)
\Theta(R^2-\eta^2-\phi^2)\,,
\end{split}\end{align}
followed by the identity
\begin{align} 
    \Theta(R^2-\eta^2-\phi^2) = 1 - \Theta(\eta^2+\phi^2 - R^2) \,.
\end{align}

We now consider the squared amplitudes for the radiation of two
collinear partons in a generic flavour channel $|A_{FF^\prime}|^2$,
which have been derived in
Refs.~\cite{Gaunt:2014cfa,Gaunt:2014xga}. Without loss of generality,
they can be expressed as
\begin{align}\label{eq:amp2dec}
|A_{FF^\prime}(k_1,k_2)|^2 = {\cal A}_{FF^\prime}^{\rm correlated}(k_1,k_2) +  {\cal A}_{FF^\prime}^{\rm uncorrelated}(k_1,k_2) \,,
\end{align}
where $ {\cal A}_{FF^\prime}^{\rm uncorrelated}$ is the contribution
that survives in the limit in which the two emissions $k_1$, $k_2$ are
infinitely far in rapidity, while the remaining part
${\cal A}_{FF^\prime}^{\rm correlated}$ encodes configurations in
which the two emissions are close in rapidity (see also
e.g.~Refs.~\cite{Gangal:2016kuo,Banfi:2014sua}).
The above decomposition is useful in that the two contributions give
rise to integrals with a different structure of rapidity divergences,
and as such they require slightly different treatments.
In the parametrisation~\eqref{eq:coordinates} each contribution to the
squared amplitude factorises as
\begin{align}
{\cal A}_{FF^\prime}^{\rm correlated/uncorrelated}(k_1,k_2) = \frac{1}{k^4_{1\perp}}\frac{1}{\zeta^2}{\cal D}_{FF^\prime}^{\rm correlated/uncorrelated} (\zeta,\eta,\phi)\,,
\end{align}
which implicitly defines
${\cal D}_{FF^\prime}^{\rm correlated/uncorrelated} (\zeta,\eta,\phi)$.
The calculation of $\Delta I_{FF^\prime}$ then involves phase-space
integrals of the type
\begin{align}
 \label{eq:MIs}
\int \frac{ \rd k_{1\perp}}{k_{1\perp}} \,\rd\eta_1
\,\frac{\rd\zeta}{\zeta}\,\rd\eta \,\frac{\rd \phi}{2\pi}\, & e^{-2
  k_{1\perp}
  \frac{e^{-\gamma_E}}{\nu}[\cosh{(\eta_1)}+\zeta\cosh{(\eta-\eta_1)}]}\,
\delta(k_1^\pm+k_2^\pm - (1-x) p^\pm)\notag\\
\times & \,{\cal
  D}_{FF^\prime} (\zeta,\eta,\phi)\, \Delta {\cal M}(p_{T}^{\rm
  veto},R^2)\,,
\end{align}
with $x$ denoting the longitudinal momentum fraction.
We parametrise the light-cone components in the delta function by
means of Eq.~\eqref{eq:coordinates} and the Sudakov parametrisation
\begin{equation}
k_i^\mu = \frac{k_i^+}{2} \bar{n}^\mu + \frac{k_i^-}{2} n^\mu + \kappa^\mu_{i,\perp}\,,
\end{equation}
where $k_i^+\equiv k_i\cdot n$, $k_i^-\equiv k_i\cdot \bar{n}$, and
$\bar{n}\cdot n = 2$. The large momentum component considered in the
argument of the delta function of Eq.~\eqref{eq:MIs} (either $+$ or
$-$) depends on whether we consider the beam functions along the
$\bar{n}^\mu=(1,0,0,1)$ or $n^\mu=(1,0,0,-1)$, respectively. Without
loss of generality, we will consider $n^\mu$ as the hard-collinear
direction, but the same considerations apply to the calculation of the
beam functions along $\bar{n}^\mu$.
In the following, we will discuss separately the treatment of the
exponential regulator for the correlated and uncorrelated
contributions to the beam function.

\paragraph{Rapidity regularisation for the correlated correction.}
The integrand of the correlated contribution vanishes by
construction in the limit of large rapidity separation. Therefore,
the only rapidity divergence in the integrals~\eqref{eq:MIs} arises
when $x=1$, namely when the rapidity of both emissions is
unconstrained.
We can handle the exponential regulator by integrating over $\eta_1$
using the delta function, and then expanding in distributions of $(1-x)$
as follows
\begin{align}
 \label{eq:regexp_correl}
\int \rd\eta_1&\,e^{-2
  k_{1\perp}
  \frac{e^{-\gamma_E}}{\nu}[\cosh{(\eta_1)}+\zeta\cosh{(\eta-\eta_1)}]}\,
\delta(k_1^-+k_2^- - (1-x) p^-) \notag\\ = & \frac{1}{p^-}
  \left[\left( - \ln\left((e^\eta +\zeta)(e^{-\eta}+\zeta)\right) +
  \ln\frac{p^-\nu}{k_{1\perp}^2} \right)\delta(1-x) +
  \frac{1}{(1-x)_+} + {\cal O}(\nu^{-1})\right]\,.
\end{align}
As usual, we kept only the leading power terms in the limit $\nu \to \infty$. 

\paragraph{Rapidity regularisation for the uncorrelated correction.}
The integrand of the uncorrelated contribution does not vanish
asymptotically in the regime of large rapidity separation. Therefore,
it features two types of rapidity divergences, which emerge in the
limits $\eta \to \pm \infty$ and $x=1$. Eq.~\eqref{eq:regexp_correl}
must be modified accordingly to deal with this more complicated
structure, and should now involve the divergence in $\eta$ as well.
We proceed by expanding in distributions also in the variable $w\equiv
e^\eta$ as follows\footnote{\label{footnote:2}
After symmetrising the integrand in 
$k_1$ and $k_2$, the integral over $w$ from 0 to $+\infty$ equals
twice the integral of $w$ from 0 to 1. 
The distributions in Eq.~\eqref{eq:regexp_uncorrel} are defined in the latter range.}
\begin{align}
 \label{eq:regexp_uncorrel}
\int \rd\eta_1& \rd\eta\,e^{-2
  k_{1\perp}
  \frac{e^{-\gamma_E}}{\nu}[\cosh{(\eta_1)}+\zeta\cosh{(\eta-\eta_1)}]}\,
\delta(k_1^-+k_2^- - (1-x) p^-) \\ = & \frac{1}{p^-}
  \int \rd w\left[ 2\left(\frac{\ln(1-x)}{1-x}\right)_+ \delta(w)-
                                             \frac{2}{(1-x)_+}\left(\ln(\zeta)
                                             - \ln\frac{p^-\nu}{k_{1\perp}^2} \right)\delta(w)\right.\notag\\
&\left.  \left( \frac{1}{(w)_+} \ln\frac{p^-\nu}{k_{1\perp}^2}-\left(\frac{\ln\left(\left(w^{-1}+\zeta\right)                                                  \left(w+\zeta\right)\right) }{w}\right)_+\right)\,\delta(1-x)\right.\notag\\
  &\left. - \left(\zeta_2 + 2 \ln(\zeta)
    \ln\frac{p^-\nu}{k_{1\perp}^2} - \ln^2\frac{p^-\nu}{k_{1\perp}^2}\right)\delta(w)\,\delta(1-x) 
   +  \frac{1}{(w)_+}\frac{1}{(1-x)_+} + {\cal O}(\nu^{-1})\right]\,.\notag
\end{align}
The integral over $\rd w $ can only be performed after inserting the
squared amplitude and the measurement function.

\paragraph{Laurent expansion in the jet radius.}
To proceed, in each of the contributions listed above, we consider the
differential equation derived by taking the derivative of the
integrals~\eqref{eq:MIs} in $R$. Since only the measurement function
depends on $R$, this amounts to the replacement
\begin{equation}
\Theta(\eta^2+\phi^2 - R^2) \to - \delta(\eta^2+\phi^2 - R^2)\,,
\end{equation}
where $R^2>\phi^2$.
The resulting integral can be evaluated as a Laurent expansion in the
jet radius $R$, that we obtain analytically in \texttt{Mathematica}
with the help of the package \texttt{PolyLogTools}~\cite{Duhr:2019tlz}.

To calculate the boundary condition, we decompose the
$\Theta(\eta^2+\phi^2 - R^2)$ function in
$\Delta {\cal M}(p_{T}^{\rm veto},R^2)$ given in
Eq.~\eqref{eq:DeltaJ-2L} as
\begin{align}\label{eq:decomposition}
\Theta \left(\eta ^2+\phi ^2-R^2\right)=\underbrace{\Theta \left(\phi ^2-R^2\right)}_{{\rm part\;}  A}+
\underbrace{\Theta \left(R^2-\phi ^2\right) \Theta \left(\eta ^2+\phi ^2-R^2\right)}_{{\rm part\;} B}\,.
\end{align}
The contribution stemming from part $A$ contains the collinear
singularity proportional to $\ln(R)$, while that arising from part $B$
is regular in the $R\to 0$ limit.
The collinear singularity in part $A$ does not directly allow us to take
the boundary condition at $R=0$. We then take an expansion by regions
around $R=R_0\ll 1$ and neglect terms of ${\cal O}(R_0^2)$.
All boundary conditions are calculated analytically with the exception
of the ${\cal O}(1)$ constant terms arising from part $B$ of
Eq.~\eqref{eq:decomposition} for the \textit{correlated} corrections,
which are obtained numerically as a grid in the $x$ variable. This is
the only non-analytic ingredient in our calculation.
We use the resulting boundary conditions to solve the differential
equation in $R$, and afterwards we take the limit $R_0\to 0$.
This procedure allows us to obtain the Laurent expansion to any order
in $R$. In this article we present results up to and including
${\cal O}(R^8)$ terms, which are sufficient to reach very high
precision in the numerical evaluation of the beam functions. Higher
order terms in $R$ could be in principle included in our
expansion.\footnote{We note in passing that an expansion in $R^2$ was
  also performed in Ref.~\cite{Gangal:2016kuo} in the context of
  rapidity-dependent jet vetoes.}

\subsection{Zero-bin subtraction and absence of soft-collinear mixing
  at two loops}
\label{sec:mixing}
The structure of SCET reproduces that of an expansion by
regions~\cite{Beneke:1997zp} of the relevant integrals occurring in
the (real and virtual) radiative corrections to the observable under
study. This method requires a full expansion of the integrals, in such
a way that any expansion of each region within the scaling
corresponding to a different region leads to scaleless
integrals.\footnote{For instance, one expects that the consistent
  expansion of a soft integral within the collinear region would
  vanish.}
In practical applications of SCET, the presence of scales related to
either additional regulators or the observable itself might render the
overlap contributions non-vanishing. This can be overcome by
subtracting by hand such overlapping regions to avoid double-counting
using the so-called zero-bin subtraction
procedure~\cite{Manohar:2006nz}.
In this sub-section we consider this procedure in the context of the
factorisation theorem in Eq.~\eqref{eq:factorisation}, more precisely in its
application to the matching coefficients
$I_{FF'}(x,Q,p_T^{\rm veto},R^2;\mu,\nu)$.
Starting from the definition of the renormalised beam function given
in Eq.~\eqref{eq:deffdiff}, we observe that
$I^{\perp}_{FF'}(x,Q,p_T^{\rm veto},R^2;\mu,\nu)$, which we extract from
Refs.~\cite{Luo:2019hmp,Luo:2019bmw}, already underwent zero-bin
subtraction and thus does not contain any overlap between soft and
collinear modes.
It is thus sufficient to discuss the new contribution
$\Delta I_{FF'}(x,Q,p_T^{\rm veto},R^2;\mu,\nu)$ computed in this
article, which still contains contamination from soft modes due to the
presence of additional scales such as the jet radius $R$ and the
rapidity regulator $\nu$.

The starting point of the zero-bin procedure is the subtraction from 
$\Delta I_{FF'}$ of its own expansions when
either one or both partons become soft. To be more precise, we
introduce the operator $\mathbb{SC}$ which acts on $\Delta I_{FF'}$ by
taking the expansion in the region in which emission $k_1$ is soft
($k_1^\mu\sim (k^+\sim \lambda,k^-\sim\lambda,k_\perp\sim\lambda)$)
and $k_2$ is collinear ($k_2^\mu\sim
(\lambda^2,1,\lambda)$).
This expansion affects both the squared amplitudes and the phase-space
constraint in the $\Delta I_{FF'}$ integrals.
Similarly, we introduce the operators $\mathbb{CS}$ and $\mathbb{SS}$
which, when acting on $\Delta I_{FF'}$,
perform the expansion of the beam function in the limit in which
$k_2$ is soft and $k_1$ is collinear or soft, respectively.
For the problem under consideration, we observe that the $\mathbb{CS}$
and $\mathbb{SS}$ operations commute, that is
$(\mathbb{CS}) (\mathbb{SS})= (\mathbb{SS}) (\mathbb{CS})$ (and
similarly for $\mathbb{SC}$).
At two loops, we can then define the zero-bin subtracted beam
functions as
\begin{equation}\label{eq:BsubtrStart}
\Delta I^{\rm subtracted}_{FF^\prime}\equiv \Delta I_{FF^\prime} -
\mathbb{CS} (\mathbb{1} - \mathbb{SS}) \Delta I_{FF^\prime} -
\mathbb{SC} (\mathbb{1} - \mathbb{SS}) \Delta I_{FF^\prime} - \mathbb{SS} \Delta I_{FF^\prime}\,,
\end{equation}
where the terms $(\mathbb{1} - \mathbb{SS})$ are
responsible for subtracting the soft-soft limit of the soft-collinear
subtraction.
For a generic channel $FF^\prime$, some of the terms in
Eq.~\eqref{eq:BsubtrStart} might vanish at leading power in the
counting parameter $\lambda$.

In this procedure, a crucial role is played by the terms
$\mathbb{CS} (\mathbb{1} - \mathbb{SS}) \Delta I_{FF^\prime} $ and
$\mathbb{SC} (\mathbb{1} - \mathbb{SS}) \Delta I_{FF^\prime} $, which
describe an interplay between the soft and collinear modes. The only
overlap between soft and collinear modes predicted by the SCET
factorisation theorem~\eqref{eq:factorisation} at a given perturbative
order amounts to products of terms arising from the
lower-order soft and beam functions. 
The absence of any other type of overlap is a necessary requirement
for the observable under consideration to factorise and therefore to
be resummable in SCET. 
In the case at hand, the terms
$\mathbb{CS} (\mathbb{1} - \mathbb{SS}) \Delta I_{FF^\prime} $ and
$\mathbb{SC} (\mathbb{1} - \mathbb{SS}) \Delta I_{FF^\prime} $ in
Eq.~\eqref{eq:BsubtrStart} are responsible for subtracting the overlap
between soft and collinear regions.
Performing the multipole expansion of the phase-space constraint is
necessary to demonstrate that these terms have the expected form and thus
that the factorisation theorem in Eq.~\eqref{eq:factorisation} is 
indeed correct.
This must be explicitly verified in the presence of the exponential
regularisation scheme, since it introduces an additional scale that
prevents integrals that would otherwise be scaleless from vanishing.

To be concrete, let us focus on the term
$\mathbb{CS} (\mathbb{1} - \mathbb{SS}) \Delta I_{FF^\prime} $ on the
r.h.s.~of Eq.~\eqref{eq:BsubtrStart}. To compute this term,
we start by acting with the operator
$\mathbb{CS} (\mathbb{1} - \mathbb{SS})$ on the measurement function
in Eq.~\eqref{eq:DeltaJ-2L}, namely on
\begin{align}\label{eq:DeltaJ-2L-full}
\Delta {\cal M}(p_{T}^{\rm veto},R^2) \equiv &\left[\Theta(p_{T}^{\rm veto}-\max\{k_{1\perp}, k_{2\perp}\})
- \Theta\left(p_{T}^{\rm veto}-\sqrt{k_{1\perp}^2+k_{2\perp}^2+2 k_{1\perp}\cdot k_{2\perp}}\right)\right]\notag\\
                                & \times \Theta(\eta^2+\phi^2 - R^2)\,.
\end{align}
We then rewrite
\begin{equation}\label{eq:clustering-shuffle}
  \Theta(\eta^2+\phi^2 - R^2) = 1 - \Theta(R^2-\eta^2-\phi^2)\,.
\end{equation}
The first term on the right-hand side of the above equation leads to a
factorising contribution, in that the contribution associated with
the theta function
$\Theta\left(p_{T}^{\rm veto}-\sqrt{k_{1\perp}^2+k_{2\perp}^2+2
    k_{1\perp}\cdot k_{2\perp}}\right)$ in
Eq.~\eqref{eq:DeltaJ-2L-full} will cancel exactly against
$I^{\perp}_{FF'}(x,Q,p_T^{\rm veto},R^2;\mu,\nu)$ when considering
the full beam function $I_{FF'}(x,Q,p_T^{\rm veto},R^2;\mu,\nu)$
as defined in Eq.~\eqref{eq:deffdiff}.
One is then left with the term
\begin{equation}
\Theta(p_{T}^{\rm veto}-\max\{k_{1\perp}, k_{2\perp}\}) = \Theta(p_{T}^{\rm veto}-k_{1\perp}) \Theta(p_{T}^{\rm veto}-k_{2\perp})\,.
\end{equation}
The trivial action of the $\mathbb{CS} (\mathbb{1} - \mathbb{SS})$
operator on this term amounts to simply replacing the transverse
momenta $k_{i\perp}$ with those of the collinear and soft
particles. The resulting phase-space integral reduces to the product
of the one loop beam and soft functions, in line with what is predicted
by the factorisation theorem~\eqref{eq:factorisation}.
Instead, the second term on the right-hand side of
Eq.~\eqref{eq:clustering-shuffle} seemingly leads to a term that is
not captured by the factorisation theorem, featuring the measurement
function
\begin{align}\label{eq:mixing}
\Delta {\cal M}^{\rm mix}(p_{T}^{\rm veto},R^2) \equiv &  \left[\Theta(p_{T}^{\rm veto}-\max\{k_{1\perp}, k_{2\perp}\})
- \Theta\left(p_{T}^{\rm veto}-\sqrt{k_{1\perp}^2+k_{2\perp}^2+2 k_{1\perp}\cdot k_{2\perp}}\right)\right]\notag\\
                                & \times\left(- \Theta( R^2-\eta^2-\phi^2 )\right)\,.
\end{align}
To proceed, we act with the $\mathbb{CS} (\mathbb{1} - \mathbb{SS})$
operator on the above measurement function. The action of
$\mathbb{CS}$ corresponds to taking the limit
$k_1^\mu\sim (\lambda^2,1,\lambda)$ and
$k_2^\mu\sim (\lambda,\lambda,\lambda)$.  Following
Ref.~\cite{Becher:2013xia}, we then expand the clustering condition in
Eq.~\eqref{eq:mixing}. Noticing that $|\eta| \gg 1$, this amounts
to~\footnote{Although at first glance Eq.~\eqref{eq:expand} resembles
  an expansion in $\ln 1/\lambda$, one could recast it as an
  equivalent power expansion in $\lambda$ as shown in section 3 of
  Ref.~\cite{Becher:2013xia}.}
 \begin{align}\label{eq:expand}
   \Theta (R^2 - \eta^2 -  \phi^2) = \Theta ( - \eta^2 ) + \delta(-  \eta^2)(R^2 - \phi^2 )+ \ldots
 \end{align}
 The phase-space constraints on the r.h.s.~of the above equation lead
 to vanishing integrals as in this region $|\eta|\gg 1$. We conclude
 that the terms 
$\mathbb{CS} (\mathbb{1} - \mathbb{SS}) \Delta I_{FF^\prime} $ and
$\mathbb{SC} (\mathbb{1} - \mathbb{SS}) \Delta I_{FF^\prime} $
vanish for the \textit{mixing}
 configuration arising from the measurement
 function~\eqref{eq:mixing}, in line with the prediction of the
 factorisation theorem~\eqref{eq:factorisation}.
 This result demonstrates that this factorisation is formally
 preserved at the two-loop order in the exponential regularisation
 scheme.
 We then perform an explicit calculation of the remaining
 non-vanishing integrals entering the definition of the zero-bin
 subtraction~\eqref{eq:BsubtrStart}.
The calculation is performed using the same approach discussed in the
previous sub-section, where now all the boundary conditions for the
$R^2$ differential equation are evaluated fully analytically. The only
technical difference with the calculation discussed in the previous
section is the treatment of the exponential regulator, which is now
modified by the fact that either one (for the soft-collinear zero bin)
or none (for the double-soft zero bin) of the momenta $k_1$, $k_2$
appear in the longitudinal $\delta$ function in the integrals
corresponding to Eq.~\eqref{eq:MIs}. The analogues of
Eqs.~\eqref{eq:regexp_correl} and~\eqref{eq:regexp_uncorrel} for these
contributions are given in Appendix~\ref{app:zerobin}.

An alternative approach to the zero-bin subtraction procedure, adopted
in Refs.~\cite{Tackmann:2012bt,Stewart:2013faa}, is to evaluate
Eq.~\eqref{eq:BsubtrStart} using an alternative set of operators
$\mathbb{\overline{CS}}$, $\mathbb{\overline{SC}}$ and
$\mathbb{\overline{SS}}=\mathbb{SS}$, where the bar indicates that
they do not act on the clustering condition
$\Theta(\eta^2+\phi^2 - R^2) $ present in the measurement function
defining $\Delta I_{FF^\prime}$, which is then left unexpanded. This
leads to the following alternative definition of the zero-bin
subtracted beam function
$\Delta \overline{I}^{\rm \,subtracted}_{FF^\prime} $:
\begin{equation}\label{eq:BsubtrStartBar}
\Delta \overline{I}^{\rm subtracted}_{FF^\prime}\equiv \Delta I_{FF^\prime} -
\mathbb{\overline{CS}} (\mathbb{1} - \mathbb{SS}) \Delta I_{FF^\prime} -
\mathbb{\overline{SC}} (\mathbb{1} - \mathbb{SS}) \Delta I_{FF^\prime} - \mathbb{SS} \Delta I_{FF^\prime}\,.
\end{equation}
Within this approach, the mixing terms originating from the integrals
given by
$\mathbb{\overline{CS}} (\mathbb{1} - \mathbb{SS}) \Delta
I_{FF^\prime} $ with the measurement function~\eqref{eq:mixing} do
not vanish any longer.
As such, to reproduce the QCD result, one has to add them
back by hand to the factorisation theorem~\eqref{eq:factorisation}.
We denote these terms by $\Delta  \overline{I}^{\rm mix}_{FF^\prime}$.
In practice, at the two loop order considered here, these are then given by
\begin{equation}
  \label{eq:mixingOps}
\Delta  \overline{I}^{\rm mix}_{FF^\prime} = \left(\mathbb{\overline{CS}}-\mathbb{CS}\right) (\mathbb{1} - \mathbb{SS}) \Delta
I_{FF^\prime} \,,
\end{equation}
where the difference $\mathbb{\overline{CS}}-\mathbb{CS}$ picks out
the contribution associated with the measurement given in
Eq.~\eqref{eq:mixing}.

Refs.~\cite{Tackmann:2012bt,Stewart:2013faa} claim that such mixing
terms constitute an ${\cal O}(R^2)$ violation of the SCET
factorisation theorem~\eqref{eq:factorisation} already at the NNLO
(and NNLL) level, making it unclear how to carry out the resummation
for the jet vetoed cross section to N$^3$LL for such ${\cal O}(R^2)$
terms.
Ref.~\cite{Stewart:2013faa} proposes to add these terms back by hand
to Eq.~\eqref{eq:factorisation} in order to achieve NNLL accuracy, but
no fix is proposed beyond this order.

One can however note that, from an explicit calculation, the
soft-collinear mixing terms at the two-loop order have the same
logarithmic structure as the zero-bin subtracted beam function
$\Delta \overline{I}^{\rm \,subtracted}_{FF^\prime}$, that is they
contain only logarithms of the type $\ln(\mu/p_{T}^{\rm veto})$ and
$\ln(\nu/p^-)$ (see also the corresponding discussion for rapidity
dependent jet vetoes in Ref.~\cite{Gangal:2016kuo}). This allows one,
at this perturbative order, to absorb them into a re-definition of the
subtracted two-loop beam functions as
\begin{align}\label{eq:BsubtrSC}
\Delta \overline{I}^{\rm \,subtracted}_{FF^\prime}  \to 
\Delta \overline{I}^{\rm \, subtracted}_{FF^\prime}  + 
2 \,\Delta \overline{I}^{\rm mix}_{FF^\prime}\,.
\end{align}
From Eq.~\eqref{eq:mixingOps} it follows that 
\begin{equation}\label{eq:mixAndNoMix}
  \Delta \overline{I}^{\rm \, subtracted}_{FF^\prime}  + 
2 \,\Delta \overline{I}^{\rm mix}_{FF^\prime}=\Delta {I}^{\rm subtracted}_{FF^\prime}\,,
\end{equation}
which we have verified by explicit calculation.
As such, we conclude that the factorisation
theorem~\eqref{eq:factorisation} is preserved at NNLO and there are no
additional mixing terms in Eq.~\eqref{eq:BsubtrSC} after performing
the multipole expansion discussed above.  The procedure leading to
Eq.~\eqref{eq:BsubtrSC} can be used as a way to compute
$\Delta I^{\rm subtracted}_{FF^\prime}$ without performing the
multipole expansion of the measurement function, but the apparent
presence of mixing terms does not constitute a breaking of the SCET
factorisation theorem at this perturbative order.
For the interested reader, together with the results for the two-loop
subtracted beam functions, we also provide in the ancillary files the
expressions for the mixing terms of Eq.~\eqref{eq:BsubtrSC},
$\Delta \overline{I}^{\rm mix}_{FF^\prime}$, obtained without
performing the multipole expansion discussed above.

We note that our findings are also consistent with the QCD formulation of the
resummation of the jet-vetoed cross
section~\cite{Banfi:2012yh,Banfi:2012jm}. Eq.~\eqref{eq:mixing}
predicts a clustering between soft
$k^\mu \sim (\lambda,\lambda,\lambda)$ and collinear
$k^\mu \sim (\lambda^2,1,\lambda)$ modes which is absent in the QCD
formulation due to the nature of the jet algorithms belonging to the
generalised $k_t$ family. By construction, these do not cluster
together partons that fly at very different rapidities, as it is the
case for a soft and a collinear parton which feature a large rapidity
separation $|\eta|\sim |\ln 1/\lambda^2 |\gg 1$. The only possible
clustering between a collinear and a soft parton in QCD is when the
latter is also collinear, i.e.~$k^\mu \sim (\lambda^2,1,\lambda)$
albeit with a small energy, which is entirely accounted for in the
definition of the beam functions.
Therefore, the absence of mixing terms in the SCET formulation is
consistent with the QCD expectation.

\subsection{Results and convergence of the small-$R$ expansion}
\label{sec:results-small-r}

The two-loop zero-bin subtracted beam functions are included as
\texttt{Mathematica}-readable files in the ancillary files
accompanying this article. 
For each channel, we decompose the result into the different colour
structures contributing at two loops. The final corrections to the
(zero-bin subtracted) matching coefficients are obtained from their
own colour decompositions as follows (we drop here the superscript
$^{\rm subtracted}$ used in the previous sub-section to simplify the
notation):
\begin{align}
  \Delta  I_{QQ}^{(2)} &= \Delta  I_{QQ_{C_F^2}}^{(2)}+\Delta
  I_{QQ_{C_AC_F}}^{(2)}+n_F \Delta  I_{QQ_{C_FT_F}}^{(2)}+\Delta
                         I_{QQ_{S}}^{(2)}\,,\notag\\
  \Delta  I_{Q\bar{Q}}^{(2)} &=\Delta
                         I_{Q\bar{Q}_{S}}^{(2)}+\Delta
                               I_{Q\bar{Q}_{NS}}^{(2)}\,,\notag\\
  \Delta  I_{Q\bar{Q}'}^{(2)} &=\Delta  I_{QQ'}^{(2)}=\Delta
                                I_{Q\bar{Q}_{S}}^{(2)}\,,\notag\\
  \Delta  I_{QG}^{(2)} &= \Delta  I_{QG_{C_AT_F}}^{(2)} +\Delta
                         I_{QG_{C_FT_F}}^{(2)} \,,\notag\\
  \Delta  I_{GQ}^{(2)} &= \Delta  I_{GQ_{C_AC_F}}^{(2)} +\Delta
                         I_{GQ_{C_F^2}}^{(2)} \,,\notag\\
   \Delta  I_{GG}^{(2)} &= \Delta  I_{GG_{C_A^2}}^{(2)}+\Delta
                          I_{GG_{C_A T_F}}^{(2)}+\Delta  I_{GG_{C_FT_F}}^{(2)}\,.
\end{align}
The full beam functions are then obtained with
Eq.~\eqref{eq:deffdiff}. We stress that in our conventions the
$\Delta I_{GG_{C_A T_F}}^{(2)}$ matching coefficients already
contain a factor of $n_F$, while the $\Delta I_{QQ_{C_FT_F}}^{(2)}$ do
not.

\begin{figure}[t]
  \begin{center}
    \includegraphics[width=0.49\columnwidth]{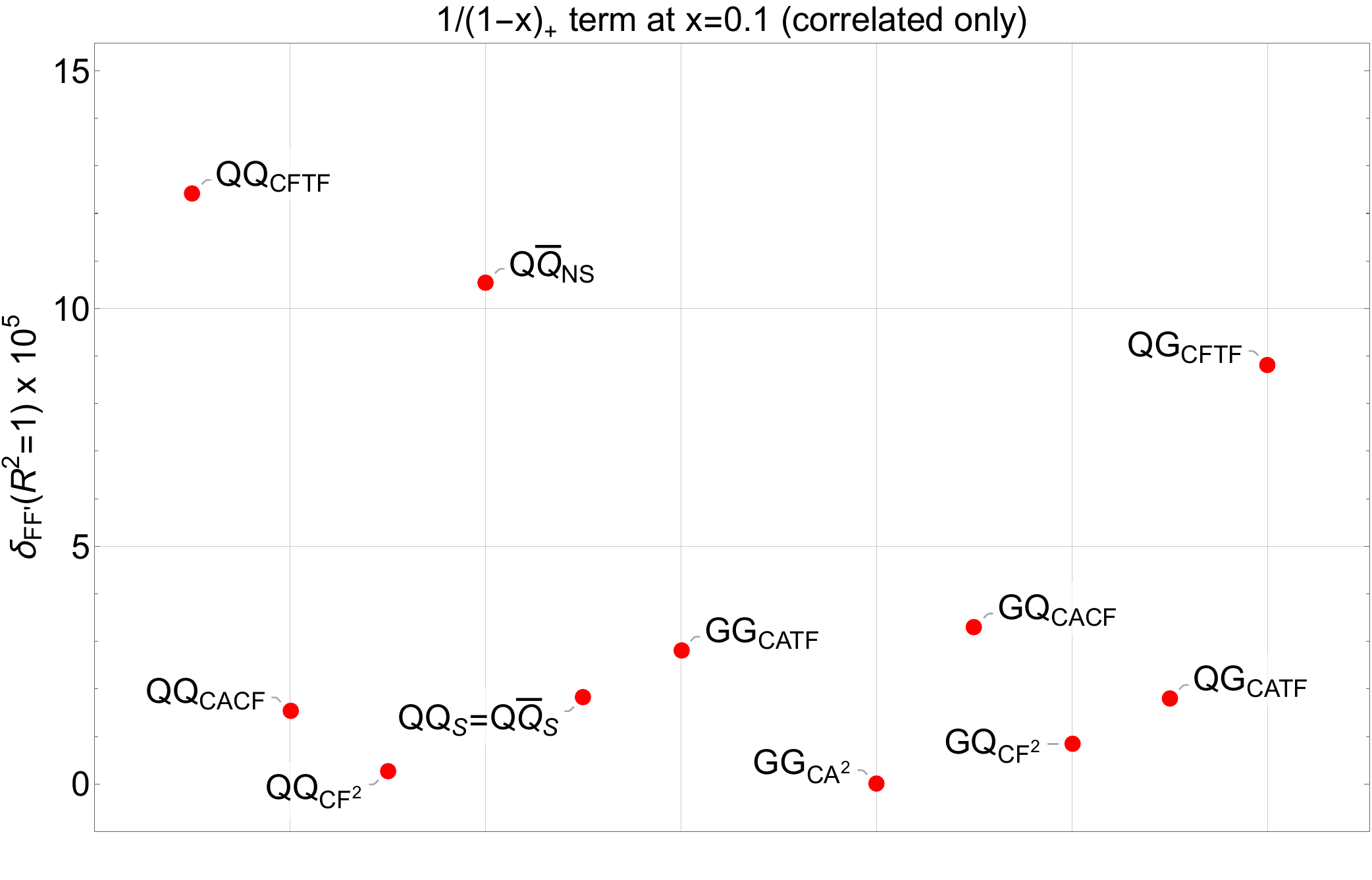}
    \includegraphics[width=0.49\columnwidth]{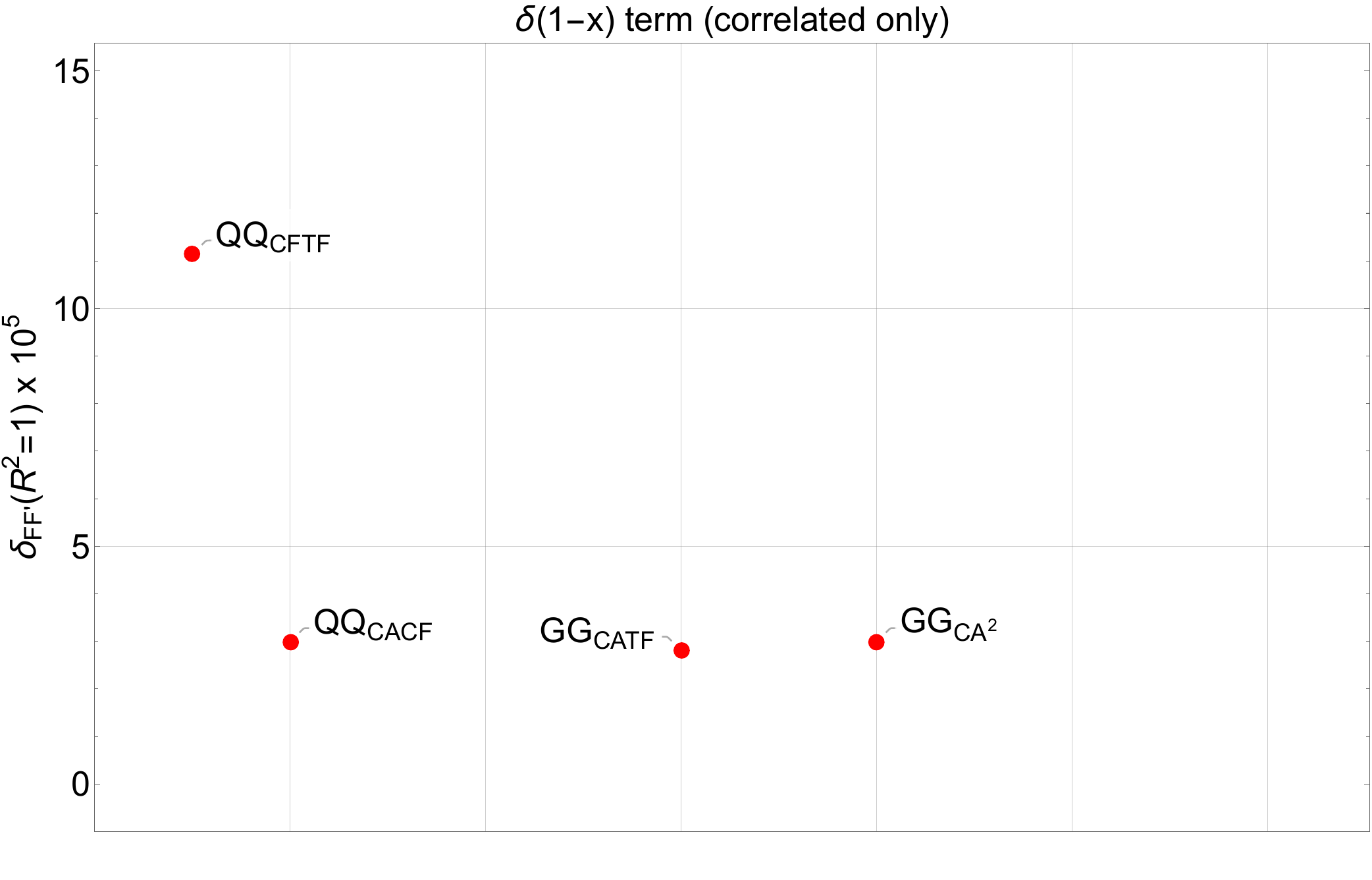}
\end{center}
\vspace{-2ex}
\caption{\label{fig:deltaR} The quantity $\delta_{FF^\prime}(R^2)$ as
  defined in Eq.~\eqref{eq:delta} evaluated at $R=1$ for the different
  flavour channels for the $1/(1-x)_+$ contribution at $x=0.1$ (left)
  and the $\delta(1-x)$ contribution (right). Note that, as indicated on the 
  $y$ axis, each value of $\delta_{FF^\prime}(R^2)$ has been
  multiplied by a factor of $10^5$.}
\end{figure}

We now discuss some consistency checks on our results and on the validity
of the small-$R$ expansion for phenomenologically relevant values of
the jet radius.
As a first check, we verified that the dependence
of the beam functions on $\ln \nu$  matches the prediction 
from the evolution equation~\eqref{eq:rapEvB}.
As a second check, to assess the validity of our expansion
in $R$ we considered the quantity
$\Delta{I}^{(2)}_{FF^\prime}$ (before performing the zero-bin
subtraction discussed in Sec.~\ref{sec:mixing}) truncated at
different orders in $R^2$. 
More precisely, we defined the relative
difference of the expansions at sixth and eighth order in $R$, 
and plotted the quantity
\begin{equation}\label{eq:delta}
  \delta_{FF^\prime}(R^2)=\left|1-\frac{\Delta I^{(2)}_{FF^\prime}(x,Q,p_{T};\mu, p_{T}^2/Q)|_{R^6}}
  {\Delta I^{(2)}_{FF^\prime}(x,Q,p_{T};\mu, p_{T}^2/Q)|_{R^8}}\right|\,,
\end{equation}
for each different flavour channel (we set
$\nu=p_{T}^2/p^-=p_{T}^2/Q$ to remove the rapidity logarithms in
$\Delta{ I}^{(2)}_{FF^\prime}$).
We find that in all cases $\delta_{FF^\prime}(R^2)$ is vanishingly small
up to $R=1$, where the convergence of the $R^2$
expansion is not necessarily guaranteed.
The convergence of the series is drastically improved at smaller values
of $R$ which are relevant for collider phenomenology.
As an example, we plot in Fig.~\ref{fig:deltaR}
the correlated corrections at $R=1$.
The plot shows an excellent convergence of the small-$R$ expansion
with residual corrections well below the permille level.

\section{Numerical computation of the quark and gluon beam functions}
\label{sec:numeric}

We now discuss a numerical evaluation of the quark and gluon beam
functions, which provides a crucial consistency check of the
analytic results discussed in the previous section. In this section,
we first outline the steps followed in the numerical calculation and 
then present a comparison between our numerical and analytic results.

\subsection{Method for the numerical computation}
All numerical integrations discussed below are performed using the
\texttt{GlobalAdaptive} \texttt{NIntegrate} routine from
\texttt{Mathematica}, to an accuracy at the permyriad level or
better. This will allow for precise numerical tests of the analytic
calculation.

\paragraph{The correlated correction.} For this part of the beam
function, the integrand can only have a rapidity divergence as
$\eta_t \equiv \tfrac{1}{2}(\eta_1 + \eta_2) \to -\infty$ at finite
$\eta \equiv \eta_1 -\eta_2$, corresponding to $x \to 1$. Note that
the restriction $k_1^- + k_2^- < p^-$ forbids us from approaching
$\eta_t \to \infty$ and encountering a rapidity divergence in this
limit.
We find it convenient to choose integration variables that remain finite as 
$\eta_t \to -\infty$ (at finite
$k_{1\perp}^2$, $k_{2\perp}^2$, $\eta$). In this way, $x$ controls the
approach to the rapidity divergence. 
Explicitly, we choose
\begin{align} \label{eq:numcorrobs}
    \mathcal{Z} \equiv \dfrac{k_1^+}{k_1^+ + k_2^+}\,, \quad \mathcal{T} \equiv (k_1^+ + k_2^+)(k_1^- + k_2^-)\,,
\end{align}
along with $\eta$ and the azimuthal separation between the two emitted partons, $\phi$.

Using these variables, the rapidity divergences  manifest
themselves as a factor of \mbox{$1/(1-x)$} in the squared amplitude. This is
regulated by inserting the exponential regulator
factor~\eqref{eq:regulator_def}, where in the exponent we may drop the
$k^- \equiv k_1^- + k_2^-$ as we do not encounter any rapidity
divergences associated with $k_1^-,k_2^- \to \infty$. We take the
limit $\nu \to +\infty$ in the regulator and make use of the
distributional expansion given in Eq.~(3.5) of \cite{Luo:2019hmp}:
\begin{align} 
\frac{e^{-\frac{1}{\nu\,(1-x)}}}{1-x} =  (\ln \nu - \gamma_E) \, \delta(1-x) + \frac{1}{(1-x)_+} + \mathcal{O}(\nu^{-1}) \, .
\label{eq:expreg_dist1}
\end{align}
The structure of the integrand in $\mathcal{T}$ is simple, containing
only terms of the form $\log^n(\mathcal{T})/\mathcal{T}$, and
integration over this variable may be done analytically. This just
leaves the integrations over $\mathcal{Z}$, $\eta$, and $\phi$, which
are performed numerically at fixed values of $x$ and $R$.

\paragraph{The uncorrelated correction.} After symmetrisation of the
integrand in partons $k_1$ and $k_2$, we may choose to integrate only
over $\eta<0$ and then multiply the final result by $2$. With this
restriction, for the uncorrelated contribution to the squared
amplitude we encounter rapidity divergences as $\eta_1 \to -\infty$,
as well as when $\eta_t \to -\infty$. For our integration variables we
should choose two variables that control the approaches to these two
limits, and then other variables that remain finite in these
limits. We choose to use:
\begin{align}
\chi_1 \equiv k_2^+ (k_1^- + k_2^-), \quad \chi_2 \equiv \dfrac{k_1^- (k_1^+ + k_2^+)^2}{k_1^+ k^2_{2\perp}} ,\quad \overline{\mathcal{Z}} = 1 - \mathcal{Z}\,,
\end{align}
as well as $x$ and $\phi$.\footnote{Note that for both this calculation and 
that of the correlated correction, an alternative convenient choice of 
variables would be those defined in Eq.~\eqref{eq:softnumvariables} as well 
as $\eta, \phi,x$.} Then, the limit $\eta_1 \to -\infty$ corresponds to
$\overline{\mathcal{Z}} \to 0$, whilst $\eta_t \to -\infty$ corresponds to 
$x \to 1$ and the rapidity divergences manifest themselves as a factor 
$1/[\overline{\mathcal{Z}}(1-x)]$ in the integrand. We insert the 
exponential regulator (again, we can drop the $k^+$ in the exponent),
and make use of the distributional expansion given in Eq.~(3.30) 
of Ref.~\cite{Luo:2019hmp}:
\begin{align} \label{eq:distexpcomplex}
\frac{1}{(1-x)\overline{\mathcal{Z}}} \exp \left( -\frac{1}{\nu(1-x)\overline{\mathcal{Z}}} \right) = \left( \frac{1}{2}(\ln \nu - \gamma_E)^2 + \frac{\pi^2}{12} \right) \delta(1-x) \, \delta(\overline{\mathcal{Z}}) + \frac{1}{(1-x)_+} \frac{1}{\overline{\mathcal{Z}}_+}
\nonumber \\
+ \left( \bigg[ \frac{\ln \overline{\mathcal{Z}}}{\overline{\mathcal{Z}}} \bigg]_+ + \frac{\ln \nu - \gamma_E}{\overline{\mathcal{Z}}_+} \right) \delta(1-x)
+ \left( \bigg[ \frac{\ln(1-x)}{1-x} \bigg]_+ + \frac{\ln \nu - \gamma_E}{(1-x)_+} \right) \delta(\overline{\mathcal{Z}})
+ \mathcal{O}(\nu^{-1}) \,.
\end{align}
We perform the integration over $\chi_1$ analytically, and the integrations over the $\chi_2$, $\overline{\mathcal{Z}}$ and $\phi$ variables numerically (for terms containing a $\delta(\overline{\mathcal{Z}})$, we perform the trivial $\overline{\mathcal{Z}}$ integration analytically).

\paragraph{The soft-collinear zero bins.} 
We use the approach of Refs.~\cite{Tackmann:2012bt,Stewart:2013faa}
as a way to compute the zero-bin subtraction without 
performing the multipole expansion of the measurement function,
see Sec.~\ref{sec:mixing} and in particular Eq.~\eqref{eq:mixAndNoMix}.
Let us, without loss of
generality, take parton $k_1$ to be soft. Then $k_1^-$ is no longer
restricted by the delta function on the minus light-cone momentum, and
we may have rapidity divergences for $\eta_1 \to \pm \infty$ as well
as for $\eta_2 \to -\infty$. We handle this calculation by
re-expressing the clustering constraint in the measurement as:
\begin{align} \label{eq:delRsimpleflip}
    \Theta(\eta^2+\phi^2 - R^2) = 1 - \Theta(R^2-\eta^2-\phi^2)\,.
\end{align}
In the second term on the right hand side, the two partons are
restricted to be close together in rapidity, such that we only have
rapidity divergences corresponding to $\eta_t \to -\infty$ (or
$x \to 1$).  The same strategy may then be used for this term as was
used for the correlated corrections. Note that there is no collinear
divergence here associated with $\eta,\phi \to 0$, due to the form of
the squared amplitude for the soft-collinear zero bin (which coincides with
the squared amplitude for the uncorrelated correction).

For the first term on the right-hand side of
Eq.~\eqref{eq:delRsimpleflip}, we choose to use the same variables
we used in Ref.~\cite{Abreu:2022sdc}:
\begin{align} \label{eq:softnumvariables}
  \mathcal{K}^2_T = k_{1\perp}^2 + k_{2\perp}^2\,,  \qquad z = \dfrac{k_{1\perp}^2}{k_{1\perp}^2 + k_{2\perp}^2}\,,
\end{align}
along with $\eta_1$, $\phi$ and $x$. The approach to
$\eta_2 \to -\infty$ is then controlled by $x$, whilst $\eta_1$
directly controls the approach to $\eta_1 \to \pm \infty$. We
introduce the exponential regulator, and split it in a straightforward
way into two factors depending on $k_1$ and $k_2$ respectively; we
drop the $k_2^+$ in the exponent as before, but now may no longer drop
$k_1^+$. The integrand does not depend on $\eta_1$ (except in the
regulator factor), and we may perform the integral over $\eta_1$ using
Eq.~(3.24) from Ref.~\cite{Abreu:2022sdc}. We utilise
Eq.~\eqref{eq:distexpcomplex} for the rapidity divergence
corresponding to $x \to 1$. The integration over $\mathcal{K}^2_T$ is
performed analytically, and the $z$ and $\phi$ integrals are done
numerically.

\paragraph{The soft-soft zero bins.} This calculation coincides
exactly with that performed in Ref.~\cite{Abreu:2022sdc} (up to a prefactor
of $\delta(1-x)$ that appears here), and we use the results presented
in that paper for this contribution.

\subsection{Comparison with the analytic results}
\label{sec:results}

\begin{figure}[t]
  \begin{center}
    \includegraphics[width=0.49\columnwidth]{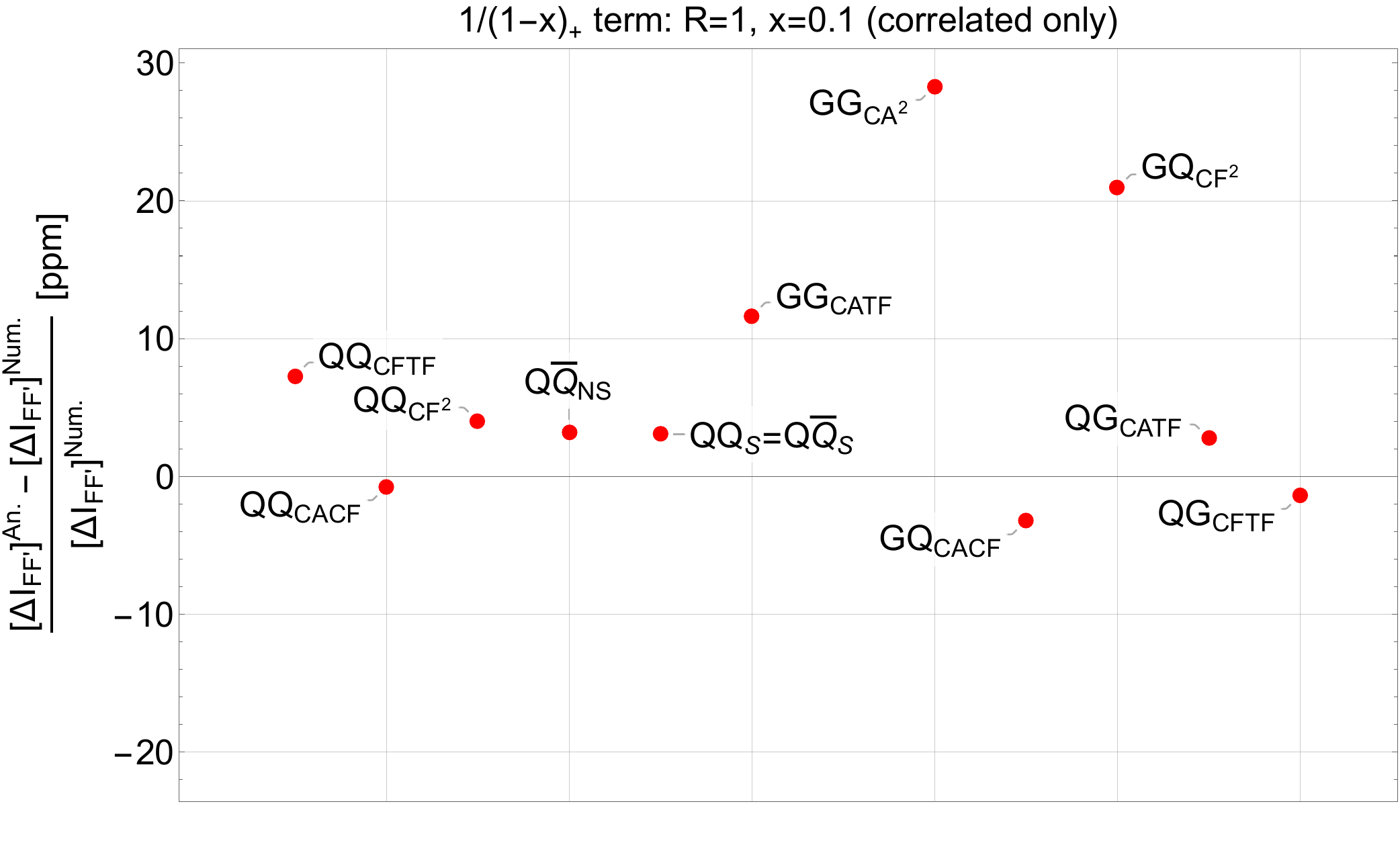}
    \includegraphics[width=0.49\columnwidth]{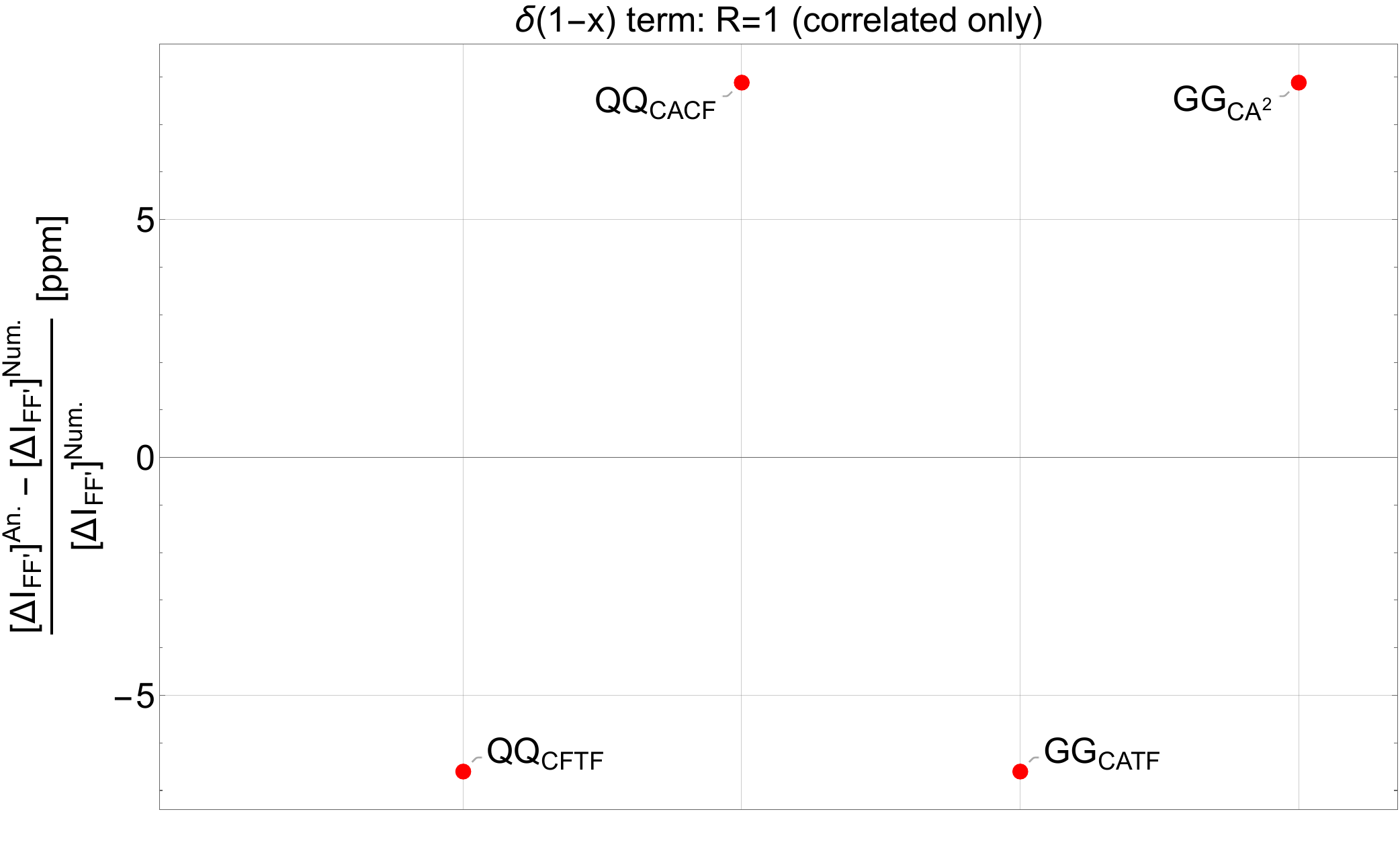}
\end{center}
\vspace{-2ex}
\caption{\label{fig:comp2num}
  Difference, in parts per million, between the analytic result and the numerical calculation with full $R$ dependence for the correlated contribution to the unsubtracted beam functions. The figures show the coefficient of $1/(1-x)_+$ at $x=0.1$ (left plot) and the coefficient of $\delta(1-x)$ (right plot), both evaluated at $R=1$. The different labels denote the various colour structures contributing to each flavour channel. The size of the difference is always at the level of the precision of the numerical calculation.}
\end{figure}

As a further assessment of the quality of our analytic small-$R$ expansion, 
we compare the numerical calculations (which have exact $R$
dependence) with the analytic results obtained in the previous section 
at different values of $R$. Due to the many flavour channels, 
we choose to show here only the
worst-case scenario, namely the comparison between the two
calculations for the most complicated contributions, corresponding to the
correlated part of the squared amplitudes in
Eq.~\eqref{eq:amp2dec}. Fig.~\ref{fig:comp2num} shows the outcome of
this comparison at $R=1$, and we can see that the difference between the
two computations is at the level of parts per million, which is the
level of accuracy of the numerical calculation.
This demonstrates that the $R$ expansion converges extremely well up to $R=1$.


\section{Leading-jet \boldmath $p_T$ slicing at NNLO}
\label{sec:slicing}

The computation of the two-loop beam functions for the leading-jet $p_T$ constitutes the last missing ingredient to construct a non-local subtraction scheme for colour singlet production at NNLO based on $p_{T}^{\rm jet}$.
In analogy with non-local subtraction schemes such as $q_T$-subtraction~\cite{Catani:2007vq}, jettiness subtraction~\cite{Boughezal:2015dva,Gaunt:2015pea}, and $k_T^{\rm ness}$ subtraction~\cite{Buonocore:2022mle}, we can formulate an NNLO slicing fully differential in the Born phase space for the production of a colour singlet $F$, as ($\rd\sigma\equiv \frac{\rd\sigma}{\rd\Phi_{\rm Born}}$)
\begin{align}\label{eq:slicingmaster}
	\rd\sigma^F_{\rm NNLO}&= \mathbb{H}^{\rm NNLO}_{\rm veto}\otimes \rd\sigma_{\rm Born} +  \lim_{p_{T,\rm cut}^{\rm jet} \rightarrow 0}  \int_{ p_{T,\rm cut}^{\rm jet}}^{+\infty} \rd p_{T}^{\rm jet} \left(\frac{\rd^2 \sigma^{F+\rm jet}_{\rm NLO}}{\rd p_{T}^{\rm jet}} - \frac{\rd^2 \sigma (p_{T}^{\rm veto})}{\rd p_{T}^{\rm veto}} \Bigg|_{p_{T}^{\rm veto}=p_{T}^{\rm jet}}^{(\as^2)}\right),
\end{align}
where the first term on the right hand side coincides with the non-logarithmic terms of the jet-veto cross section Eq.~\eqref{eq:factorisation}, the last term is its derivative with respect to $p_{T}^{\rm veto}$ expanded through ${\cal O}(\as^2)$ relative to the Born, and the second term is the NLO cross-section for the production of the colour singlet in association with a jet.
The above formula formally reduces to the NNLO result in the limit \mbox{$p_{T,\rm cut}^{\rm jet}~\rightarrow~0$}. 
However, since the second and the third terms are both divergent logarithmically in this limit, Eq.~\eqref{eq:slicingmaster} can be computed numerically only by choosing a finite value of $p_{T,\rm cut}^{\rm jet}~>~0$.
This introduces a slicing error $\mathcal O( (p_{T,\rm cut}^{\rm jet}/Q)^m)$, where $m$ is an integer value to be determined by studying the $p_{T,\rm cut}^{\rm jet} \rightarrow 0$ behaviour of the \emph{non-singular} contribution contained within brackets in Eq.~\eqref{eq:slicingmaster}.

\begin{figure}[t]
  \begin{center}
  \includegraphics[width=0.49\columnwidth]{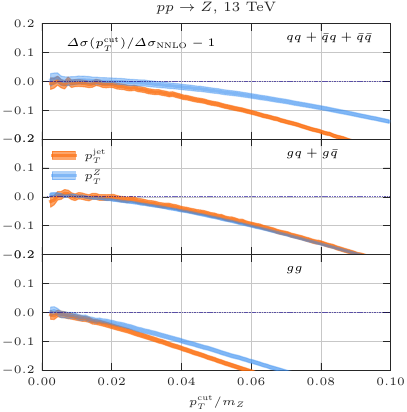}
  \includegraphics[width=0.49\columnwidth]{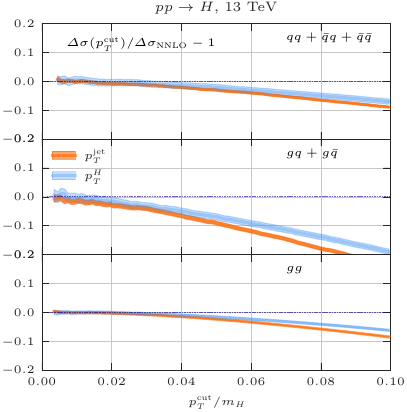}
\end{center}
\vspace{-2ex}
\caption{\label{fig:slicing}
  NNLO correction for $Z$ (left panel) and $H$ (right panel) production: leading-jet $p_T$ subtraction (blue) against $q_T$ subtraction (blue) and analytic results.}
\end{figure}

The comparison of the NNLO results obtained using Eq.~\eqref{eq:slicingmaster} to the known NNLO cross sections provides a very robust check of the correctness of the results presented in this work.
We perform this test by considering on-shell $Z$ and $H$ production, which allows us to independently check the quark and gluon beam functions, respectively.
We use \texttt{MCFM 9.1}~\cite{Campbell:2019dru} to compute the NLO result for $Z+j$~\cite{Giele:1993dj} and $H+j$~\cite{deFlorian:1999zd,Ravindran:2002dc,Glosser:2002gm} production, while we use the implementation of the jet-veto resummation~\cite{Becher:2012qa,Banfi:2012jm,Stewart:2013faa,Becher:2013xia} in the \texttt{RadISH} code~\cite{Monni:2016ktx,Bizon:2017rah,Monni:2019yyr} to compute the factorised expression~\eqref{eq:factorisation} and its expansion up to NNLO.
We compare our results with the analytic NNLO cross section for $Z$~\cite{Hamberg:1990np,vanNeerven:1991gh} and $H$~\cite{Ravindran:2002dc,Ravindran:2003um,Harlander:2002wh} production which we computed using the \texttt{n3loxs} code.\footnote{We are grateful to the authors of the \texttt{n3loxs} code for providing a preliminary version of the code to carry out our numerical checks.}
For our numerical checks, we consider proton-proton collisions at a centre-of-mass energy of 13 TeV and $R=0.4$.
We adopt the \texttt{LUXqed\_plus\_PDF4LHC15\_nnlo\_100} parton distribution functions~\cite{Manohar:2017eqh} through the \texttt{LHAPDF} interface~\cite{Buckley:2014ana}.
We choose factorisation and resummation scales equal to $\mu_R=\mu_F=m_Z,m_H$ for $Z$ and $H$ production, respectively, with $m_Z=91.1876$ and $m_H=125$~GeV.
In Fig.~\ref{fig:slicing} we study the dependence of the NNLO correction on $p_{T, \rm cut}^{\rm jet}/Q$ for $Z$ and $H$ production for different partonic channels by normalising it to the analytic result.
We compare the results obtained using $p_{T}^{\rm jet}$-slicing (in orange) with those obtained using $q_T$-slicing (in blue) to assess the performance of the two methods.
For $Z$ production we are able to lower the value of $p_{T,\rm cut}^{\rm jet}$ down to $0.1$ GeV, whereas we stop at $p_{T,\rm cut}^{\rm jet}=0.5$ GeV for Higgs production as the fixed order $H+j$ calculation becomes slightly unstable in some channels below this value.\footnote{We thank A.~Huss for providing results calculated with the \textsc{NNLOJET} code~\cite{Chen:2016zka} at $p_{T,\rm cut}^{\rm jet}=0.1$~GeV for Higgs production, which we used as an independent cross-check.}
We observe that in all the channels the results obtained using leading-jet $p_T$ slicing converge to the exact cross section in the $p_{T, \rm cut}^{\rm jet} \rightarrow 0$ limit, thus providing a powerful check of the validity of our computations.
By comparing the results obtained with $p_{T}^{\rm jet}$-slicing to those obtained using $q_T$-slicing we notice that the convergence towards the analytic result is comparable between the two methods, with $q_T$-slicing converging slightly faster in most cases for $R=0.4$. Smaller values of the jet radius $R$ appear to improve the convergence of the $p_{T}^{\rm jet}$ subtraction, possibly due to the reduced size of the subleading power corrections.
Further investigations on the size of subleading power corrections deserve dedicated studies.

\section{Conclusions}
\label{sec:conclusions}
In this article, we presented the first calculation of the complete set of two-loop beam functions relevant for the leading-jet transverse momentum resummation in colour singlet production. The results were obtained using two independent methods: a semi-analytical expansion for small jet-radius $R$ up to and including terms of $\mathcal{O}(R^8)$, and a fully numerical evaluation for several fixed values of $R$. The small-$R$ expansion is analytical with the only exception being a set of $R$-independent regular terms. The numerical calculation retains the complete $R$ dependence and shows perfect agreement with the analytical expansion in the range $R\in [0,1]$ which is relevant for collider phenomenology.
We further checked our computation by performing an NNLO calculation of the total cross section for Higgs and $Z$ boson production using a slicing subtraction scheme based on the leading-jet $p_T$. Our calculation reproduces known analytic predictions for the NNLO total cross section in all flavour channels, thus validating our results.

When describing the technical aspects of the calculation, we discussed in detail the complications related to zero-bin subtraction and soft-collinear mixing. In particular, we explicitly showed that if one performs a multipole expansion of the measurement functions
there exist no mixed soft-collinear contributions which break the SCET factorisation theorem at NNLO.
This observation is non-trivial in the presence of the exponential rapidity regulator in that it adds a new scale to the problem, which leads to the presence of non-vanishing integrals that would otherwise be scaleless.

Our complete results are provided in \texttt{Mathematica}-readable files attached to the \texttt{arXiv} version of this article. Together with our earlier analytic results for the leading-jet $p_T$ soft function \cite{Abreu:2022sdc}, they constitute a critical component of the N$^3$LL resummation of this observable, with the only missing ingredient being the three-loop rapidity anomalous dimension.

\acknowledgments
We are grateful to Thomas Becher for helpful discussions on the cancellation of soft-collinear mixing terms in SCET factorisation.
We would also like to thank Julien Baglio, Claude Duhr and Bernhard Mistlberger for providing us with a preliminary version of their computer code~\texttt{n3loxs} used in our checks of the total cross section, and Alexander Huss for kindly providing a cross check of the differential distributions with the \textsc{NNLOJet} code.
The work of JRG is supported by the Royal Society through Grant
URF\textbackslash R1\textbackslash 201500.
LR has received funding from the Swiss National Science Foundation (SNF) under contract
PZ00P2$\_$201878.
RS is supported by the United States Department of Energy under Grant
Contract DE-SC0012704.

\paragraph{Note added} In the final stages of the preparation of this article,
Ref.~\cite{Bell:2022nrj} appeared with a numerical calculation of the beam 
functions in the quark channel. These results are obtained with a different 
rapidity regulator and computed for a discrete set of real points in the 
Mellin variable $N$ conjugate to the longitudinal momentum $x$. For this reason,
it is not immediately clear how to compare the results of 
Ref.~\cite{Bell:2022nrj} with the ones  presented here.

\appendix

\section{Expansion of the exponential regulator in zero-bin integrals}
\label{app:zerobin}
In this appendix we provide the ingredients to calculate the integrals contributing to the zero-bin subtraction discussed in Sec.~\ref{sec:mixing}.
Specifically, we provide the analogues of
Eqs.~\eqref{eq:regexp_correl},~\eqref{eq:regexp_uncorrel} needed for
the calculation of the correlated and uncorrelated contributions,
respectively.

\paragraph{Soft-collinear zero-bin.} 
We consider the limit in which one of the two partons is soft (say $k_2$) and the second is collinear. The exponential regulator in the correlated corrections can be expanded as:
\begin{align}
 \label{eq:regexp_correl_sc}
\int \rd\eta_1&\,e^{-2
  k_{1\perp}
  \frac{e^{-\gamma_E}}{\nu}[\cosh{(\eta_1)}+\zeta\cosh{(\eta-\eta_1)}]}\,
\delta(k_1^-- (1-x) p^-) \notag\\ = & \frac{1}{p^-}
  \left[\left( - \ln\left(1+e^{-\eta}\zeta\right) +
  \ln\frac{p^-\nu}{k_{1\perp}^2} \right)\delta(1-x) +
  \frac{1}{(1-x)_+} + {\cal O}(\nu^{-1})\right]\,.
\end{align}
Similarly, we can use the following formula to deal with the 
uncorrelated contribution (see footnote \ref{footnote:2}):
\begin{align}
 \label{eq:regexp_uncorrel_sc}
\int \rd\eta_1& \rd\eta\,e^{-2
  k_{1\perp}
  \frac{e^{-\gamma_E}}{\nu}[\cosh{(\eta_1)}+\zeta\cosh{(\eta-\eta_1)}]}\,
\delta(k_1^-- (1-x) p^-) \\ = & \frac{1}{p^-}
  \int \rd w\left[\frac{1}{(w)_+}\frac{1}{(1-x)_+} -
                                             \frac{2}{(1-x)_+}\left(\ln(\zeta)
                                             - \ln\left(\frac{\nu}{k_{1\perp}}\right) \right)\delta(w)\right.\notag\\
&\left.   \left[-\left(\frac{\ln\left(1+w^{-1}\zeta\right)}{w}\right)_+ + \ln\left(\frac{p^- \nu }{k^2_{1\perp}} \right)\frac{1}{(w)_+}\right]\,\delta(1-x)\right.
\notag\\
  &\left. - 2\left(\ln(\zeta)-
    \ln\left(\frac{\nu}{k_{1\perp}}\right)\right) \ln\left(\frac{p^- \nu}{k^2_{1\perp}}\right)  \delta(w)\,\delta(1-x) + {\cal O}(\nu^{-1})\right]\,.\notag
\end{align}
Analogous expressions hold for the case in which $k_1$ is soft.

\paragraph{Double-soft zero-bin.} 
In the limit in which both partons are soft, the exponential regulator in the correlated corrections can be expanded as:
\begin{align}
 \label{eq:regexp_correl_ss}
\int \rd\eta_1&\,e^{-2
  k_{1\perp}
  \frac{e^{-\gamma_E}}{\nu}[\cosh{(\eta_1)}+\zeta\cosh{(\eta-\eta_1)}]}\,
\delta((1-x) p^-) \notag\\ = & \frac{1}{p^-}
  \left[\left( - \ln\left((e^\eta +\zeta)(e^{-\eta}+\zeta)\right) +
  \ln\frac{\nu^2}{k_{1\perp}^2} \right)\delta(1-x) + {\cal O}(\nu^{-1})\right]\,,
\end{align}
and in the uncorrelated correction as (see footnote \ref{footnote:2}):
\begin{align}
 \label{eq:regexp_uncorrel_ss}
\int \rd\eta_1& \rd\eta\,e^{-2
  k_{1\perp}
  \frac{e^{-\gamma_E}}{\nu}[\cosh{(\eta_1)}+\zeta\cosh{(\eta-\eta_1)}]}\,
\delta((1-x) p^-) \\ = & \frac{1}{p^-}
  \int \rd w\left[ -
                         \left[\left(\frac{\ln\left(\left(w^{-1}+\zeta\right)\left(w+\zeta\right)\right)}{w}\right)_+
                         - 2\ln\left(\frac{\nu}{k_{1\perp}}\right) \frac{1}{(w)_+}\right]\,\delta(1-x)\right.\notag\\
  &\left. - 4\ln\left(\frac{\nu}{k_{1\perp}}\right)\left(\ln(\zeta) -
    \ln\left(\frac{\nu}{k_{1\perp}}\right)\right)\delta(w)\,\delta(1-x)+ {\cal O}(\nu^{-1})\right]\,.\notag
\end{align}

\section{Renormalisation of the beam functions in SCET$_{\rm II}$}
\label{app:renorm}
In this appendix we show how the renormalisation of the 
transverse-momentum and jet-veto matching coefficients
determines the renormalisation of $\Delta I$
in Eq.~\eqref{eq:deffdiff}.
We start by writing Eq.~\eqref{eq:deffdiff} for bare quantities
\begin{align}\label{eq:bare}
I_{\rm bare}(x,Q,p_T^{\rm veto},R^2;\mu,\nu) = I^\perp_{\rm bare}(x,Q,p_T^{\rm veto};\mu,\nu)+\Delta I_{\rm bare}(x,Q,p_T^{\rm veto},R^2;\mu,\nu)\; .
\end{align}
The previous equation can be simply thought of as a way
to decompose $I_{\rm bare}$ into the sum of two
terms.
We now relate Eq.~\eqref{eq:bare} to Eq.~\eqref{eq:deffdiff}
by considering the renormalisation of the UV poles in
$\epsilon$. 

An important subtlety here is that the matching coefficient (or beam
functions) for transverse momentum and jet veto resummation
renormalise in a multiplicative way in different spaces, that is in
impact-parameter and momentum (cumulant) space, respectively. We then
use
\begin{align}\begin{split}\label{eq:renormZ}
  I &= Z\, Z_{\alpha_s}\,I_{\rm bare}\,,\\
  I^\perp&= Z^\perp \otimes \left(Z_{\alpha_s} \,I^\perp_{\rm bare} \right)\,,
\end{split}\end{align}
where the matching coefficients in the l.h.s.\ are renormalised, $Z$ is
the UV renormalisation constant, and $Z_{\alpha_s}$ accounts for the
renormalisation of the coupling in the $\overline{\textrm{MS}}$ scheme
(which can be performed in either
space). The convolution operator denoted by $\otimes$ reduces to a
simple product in impact-parameter space.
Since we are working in SCET$_{\rm II}$, the UV renormalisation
constants $Z$ and $Z^\perp$ are the same as a consequence of the fact
that the two observables have the same $\mu$ anomalous dimensions
(while the rapidity renormalisation groups structure differs), that is
$Z=Z^\perp$.
This implies that they do not depend on the value of the observable
itself (i.e.\ on $p_T^{\rm veto}$), but rather on $\mu$ and $\nu$ only
which we keep generic.
By relating Eqs.~\eqref{eq:bare} and~\eqref{eq:deffdiff} we can express
the renormalised $\Delta I$ in terms of bare quantities.
At the one-loop level one has
$I^{(1)}_{\rm bare}=I^{\perp,\, (1)}_{\rm bare}$, and hence
$\Delta I^{(1)}=\Delta I^{(1)}_{\rm bare}=0$.
At the two-loop level, since the renormalisation constants are
independent of the observable value, the convolution in
Eq.~\eqref{eq:renormZ} becomes a product leading to
\begin{equation}\label{eq:renorm2l}
\Delta I^{(2)} = \Delta \overline{I}^{ (2)}_{\rm bare} \,,
\end{equation}
where $ \Delta \overline{I}^{(2)}_{\rm bare}$ denotes the two-loop bare
matching coefficient whose coupling constant has been renormalised in
the $\overline{\rm MS}$ scheme. Eq.~\eqref{eq:renorm2l} 
justifies Eq.~\eqref{eq:deffdiff}, which was
 used in Sec.~\ref{sec:computation} for our
computation. Additional terms due to renormalisation contribute to
$\Delta I$ at higher loop orders.

We note that some of the arguments used to arrive at
Eq.~\eqref{eq:renorm2l} do not apply in the SCET$_{\rm I}$ case, where the $\mu$
anomalous dimension is observable dependent and therefore the convolution structure
already plays a role at two loops. An analogous discussion
in the  SCET$_{\rm I}$ case was discussed in Ref.~\cite{Gangal:2016kuo}.

\bibliographystyle{JHEP} \bibliography{jet}
\end{document}